\RequirePackage{fix-cm}
\documentclass[smallextended,natbib]{svjour3}
\usepackage{xcolor}
\usepackage{color}
\usepackage{siunitx}
\usepackage{textcomp}
\usepackage{framed}
\usepackage{tcolorbox}
\usepackage{booktabs}
\usepackage{algorithmic}
\usepackage{algorithm}
\usepackage{textcomp}
\usepackage{enumitem}
\usepackage{amsmath,amsfonts,amssymb}
\usepackage{listings}
\usepackage{microtype}
\usepackage{multirow}
\usepackage{graphicx}
\usepackage{threeparttable}
\usepackage{tablefootnote}
\usepackage{subfigure}
\usepackage{caption}
\usepackage{longtable}
\usepackage{soul}
\usepackage{float}
\usepackage{makecell}
\usepackage{url}
\usepackage{lineno}
\usepackage[misc]{ifsym}

\lstdefinestyle{JavaStyle}{
    language=Java,
    basicstyle=\footnotesize\ttfamily,
    keywordstyle=\color{blue},
    commentstyle=\color{green!70!black},
    stringstyle=\color{orange},
    numbers=left,
    numberstyle=\tiny\color{gray},
    stepnumber=1,
    numbersep=5pt,
    backgroundcolor=\color{gray!5},
    frame=lines,
    rulecolor=\color{black},
    breaklines=true,
    showstringspaces=false,
    tabsize=2,
    captionpos=b
    escapeinside={(*@}{@*)},
}
\smartqed
\begin{document}
\begin{sloppypar}

\title{Deep Learning Based Identification of Inconsistent Method Names: How Far Are We?}
\titlerunning{Empirical Study on DL-based Identification of Inconsistent Method Names.}

\author{Taiming Wang \textsuperscript{1}     \and
        Yuxia Zhang  \textsuperscript{1,*} \and
        Lin Jiang \textsuperscript{2}  \and
        Yi Tang \textsuperscript{3} \and
        Guangjie Li \textsuperscript{3} \and
        Hui Liu \textsuperscript{1,*}
}

\institute{
\Letter Yuxia Zhang\\
\email{yuxiazh@bit.edu.cn}\\  
\Letter Hui Liu\\
\email{liuhui08@bit.edu.cn}\\
\at
 {1} School of Computer Science \& Technology, Beijing Institute of Technology, Beijing, China.
 \email{wangtaiming@bit.edu.cn, yuxiazh@bit.edu.cn, liuhui08@bit.edu.cn}
 \at
 {2} China Telecom Corporation Limited Beijing Research Institute, Beijing, China.
  \email{jiangl34@chinatelecom.cn}
 \at
 {3} Academy of Military Science of the People's Liberation Army National Innovation Institute of Defense Technology, Beijing, China.
 \email{xtangee@hotmail.com, liguangjie\_er@126.com}
}



\date{Received: date / Accepted: date}

\maketitle

\begin{abstract}
For any software system, concise and meaningful method names are critical for program comprehension and maintenance. However, for various reasons, the method names might be inconsistent with their corresponding implementations. Such inconsistent method names are confusing and misleading, often resulting in incorrect method invocations. To this end, a few intelligent deep learning-based approaches based on neural networks have been proposed to identify such inconsistent method names in the industry. Existing evaluations suggest that the performance of such DL-based approaches is promising. 
However, the evaluations are conducted with a perfectly balanced dataset where the number of inconsistent method names is exactly equivalent to that of consistent ones. In addition, the construction method of this balanced dataset is flawed, leading to false positives in this dataset. Consequently, the reported performance may not represent their efficiency in the field where most method names are consistent with their corresponding method bodies and only a small part of method names are inconsistent with corresponding method bodies. To this end, in this paper, we conduct an empirical study to assess the state-of-the-art DL-based approaches in the automated identification of inconsistent method names.
We first build a new benchmark (dataset) by using both automatic identification from commit history and manual inspection by developers, aiming to reduce the number of false positives.  Based on the benchmark, we evaluate five representative DL-based approaches to identifying inconsistent method names (one is retrieval-based and four are generation-based). Our evaluation results suggest that the performance of the evaluated approaches is substantially reduced when we switch from the existing balanced dataset to our new benchmark.  Furthermore, to reveal where and why the evaluated approaches work/fail, we conduct quantitative and qualitative analyses of the evaluation results. Our analysis results suggest that the evaluated approaches work well on methods with simple bodies and short names, and retrieval-based approaches are especially good at methods whose names start with popular first sub-tokens.
Retrieval-based approaches fail frequently because the adopted method representation technique is not efficient enough. Another possible reason for the failures is their unverified rationale, i.e., two methods with similar bodies should have similar names. Generation-based approaches frequently fail because of inaccurate similarity calculation formulas and immature method name generation techniques. Through the data analysis, we also propose two possible ways for better identifying inconsistent method names by leveraging contrastive learning and LLMs. Overall, our empirical study suggests that the state-of-the-art DL-based approaches in inconsistent method name identification deserve significant improvement before applying them to practical software systems.
\end{abstract}

\section{Introduction}
\label{sec:Introduction}
Identifiers, i.e., the names of software entities, make up approximately 70\% of source code~\citep{Deissenboeck2006}. Such identifiers play an important role in program comprehension and software maintenance~\citep{Allamanis2015,butler2010exploring,8930886}. Therefore, their quality is critical~\citep{Lawrie2006,Schankin2018,8090141,DBLP:journals/ese/BinkleyDLMMS13,7097720}. As an important type of identifier, concise and meaningful method names can provide intuitive information about the method behaviors~\citep{Allamanis2016,DBLP:conf/icse/AlsuhaibaniNDCM21}. Developers often guess the functionalities of methods according to the short and meaningful method names instead of complex and lengthy implementations (method bodies)~\citep{gethers2011codetopics,bavota2013methodbook,deissenboeck2015concise}. However, naming a method properly is not an easy case for developers. Studies suggest that naming software entities is one of the most difficult tasks for programmers~\citep{johnson2018arg,johnson2018don}. As a result, developers often write poor (i.e., inconsistent) method names in programs due to a lack of thesaurus, conflicting styles during collaboration, and improper code cloning~\citep{kim2016automatic}.
The improper (inaccurate) method names tend to lead misunderstandings~\citep{takang1996effects,DBLP:conf/ppig/LiblitBS06,DBLP:journals/ese/ArnaoudovaPA16,hofmeister2017shorter,DBLP:conf/wcre/Arnaoudova10} and may result in incorrect method invocation and software defects~\citep{butler2009relating,amann2018systematic,abebe2011effect,abebe2012can,DBLP:conf/icsm/AghajaniNBL18}. Furthermore, these naming issues could negatively affect other software applications that rely on them~\citep{Allamanis2015}. 
Many companies are paying attention to naming conventions and coding standards~\citep{Allamanis2014}. Identifying inconsistent method names in software projects could significantly improve the quality of the codebase and lower the cost of project maintenance~\citep{butler2009relating}.
Since consistency in software engineering is well-understood to be consistent usage of naming style, indentation, etc., inconsistency may exist between sets at different levels, e.g., a collection of method names or just a method with both name and body. To clarify the meaning of inconsistency and avoid misunderstanding, here we define "inconsistent method names":
An inconsistent method name is a method name that cannot fully deliver the semantic meaning of its corresponding method body, i.e., inconsistent with its body, which could be misleading and result in possible software defects.

A few DL-based approaches have been proposed to identify inconsistent method names~\citep{host2009debugging,Liu2019,Nguyen2020,Li2021}. H{\o}st et al.~\citep{host2009debugging} build the first mining-based approach that is specially designed to identify inconsistent method names. With the increasing popularity of deep learning techniques and the principle of software naturalness~\citep{DBLP:journals/cacm/HindleBGS16,DBLP:journals/csur/AllamanisBDS18,DBLP:conf/wcre/0008NBL19,DBLP:conf/iwpc/ArimaHK18,DBLP:conf/icse/RayHGTBD16}, DL-based approaches have been proposed to identify inconsistent method names~\citep{Liu2019,Nguyen2020,Li2021}. Liu et al.~\citep{Liu2019} proposed an approach (we noted it as $IRMCC$ for convenience in the rest of the paper) that debugs inconsistent method names by leveraging convolutional neural network (CNN)~\citep{DBLP:journals/nn/MatsuguMMK03}, Word2vec~\citep{DBLP:conf/nips/MikolovSCCD13}, and Paragraph2vec~\citep{DBLP:conf/icml/LeM14}. It is the first approach that combines deep learning techniques with information retrieval (IR) techniques to solve the problem of inconsistent method name identification and yields good performance. $MN_{IRE}$ proposed by Nguyen et al.~\citep{Nguyen2020} is another DL-based approach. It first generates a name for the target method and then determines whether the name is inconsistent by comparing the similarity between the generated name and the original one. $MN_{IRE}$ is the first generation-based approach in the automated identification of inconsistent method names, and its performance is even better than that of $IRMCC$.  
Li et al.~\citep{NameChecker} proposed to identify inconsistent method names using a DL-based classifier.
\emph{DeepName}, proposed by Li et al.~\citep{Li2021}, and \emph{Cognac}, proposed by Wang et al.~\citep{wang2021lightweight}, leverage the same strategy as $MN_{IRE}$ but incorporate more types of context to identify inconsistent method names. Overall, the DL-based approaches have obtained a promising performance.

However, despite the promising performance as reported, we still lack a comprehensive understanding of the state-of-the-art DL-based approaches in the automated identification of inconsistent method names.
We notice that all the existing evaluations employed a single testing dataset where the number of inconsistent method names exactly equals that of consistent ones. However, testing data in the field is significantly imbalanced because most method names (especially those in high-quality projects) are consistent with their corresponding method bodies~\citep{Liu2019}. As a result, the number of inconsistent method names is significantly smaller than that of consistent ones. It remains unknown whether replacing the existing perfectly balanced testing dataset with a seriously imbalanced new dataset will result in changes in the performance of the evaluated approaches. In addition, the construction method of Liu et al. is flawed because it cannot guarantee that the method name changes are associated with the inconsistency between names and bodies~\citep{Wen2020, Wen2022, Kim2023}, leading to false positives in this dataset. Therefore, a new benchmark is needed to evaluate the performance of DL-based approaches in real-world scenarios and also to facilitate the evaluations of subsequent approaches. Besides the unclear overall performance of the state-of-the-art DL-based approaches on more realistic testing datasets, it also remains unknown where and why such DL-based approaches work or fail. These open questions are important for both researchers and practitioners: They should know whether the state-of-the-art DL-based approaches are accurate enough to be applied in the field, and if not, how to improve them.

To this end, in this paper, to provide some practical guidelines in the future development of DL-based approaches, we conduct an extensive empirical study on the state-of-the-art DL-based approaches in the identification of inconsistent method names on a large-scale test dataset ($BenMark$) including 2,443 inconsistent methods and 1,296,743 consistent ones. The new dataset is constructed over 430 high-quality projects collected by Liu et al.~\citep{Liu2019}. We selected five representative DL-based approaches, i.e., the approach proposed by Allamanis et al.~\citep{Allamanis2016} (noted as $CAN$), the approach proposed by Liu et al.~\citep{Liu2019} (noted as $IRMCC$), $MN_{IRE}$~\citep{Nguyen2020}, $Cognac$~\citep{wang2021lightweight}, and $GTNM$~\citep{liu2022learning} for the empirical study. 

Our empirical study investigates the following research questions:
\begin{itemize}
    \item \textbf{RQ1: How do the selected DL-based approaches perform when switching from within-project setting to cross-project setting?} \\
    \textbf{Motivation}: To the best of our knowledge, all the existing approaches are evaluated in a within-project setting. However, although within-project can be useful in some cases, it requires repetitious training of the models, which is often time-consuming and resource-consuming. By contrast, in the cross-project settings, users can train the model once and use this model to predict any new test data, which significantly facilitates usage. Therefore, we design this research question to investigate the application of existing approaches in the popular cross-project setting.  Answering RQ1 would reveal whether the state-of-the-art DL-based approaches can be used with cross-project settings and also achieve promising performance.\\
    \textbf{Answer:} Switching from a within-project setting to a cross-project setting slightly changes the performance of the evaluated approaches. The evaluated approaches can achieve promising performance in both cross-project settings and within-project settings. Since cross-project pattern requires no repeated training, such approaches could be used (by their potential users) more conveniently without a substantial reduction in performance.

    \item \textbf{RQ2: How do the selected DL-based approaches perform when changing the ratio of inconsistent and consistent method names in testing data?} \\
    \textbf{Motivation}: Note that all the existing approaches are evaluated in a dataset where the number of inconsistent method names exactly equals that of consistent ones. However, consistent method names are more popular than inconsistent ones in the field. Therefore, we design this research question to investigate whether the existing approaches can still achieve promising performance in a more realistic setting. Answering RQ2 would reveal how the evaluated approaches will be impacted by the ratio of inconsistent and consistent methods in testing data.\\
    \textbf{Answer:} The performance of the evaluated approaches is impacted substantially by the ratio of inconsistent and consistent method names.  Lowering the ratio of inconsistent method names results in a dramatic decline in the precision of identifying inconsistent method names. Existing DL-based approaches for method name consistency checking may
not work accurately in the field.

    \item\textbf{RQ3: Where and why do the IR-based approaches work or fail?}\\
    \textbf{Motivation}: Since few related approaches have been proposed since 2022, an empirical study is desperately needed to motivate the design of more advanced approaches. Answering RQ3 would reveal the strengths and weaknesses of IR-based approaches and may provide insights for further research aimed at improving their performance.\\
    \textbf{Answer:} IR-based approach, i.e., $IRMCC$ works better on methods with simple bodies and methods whose names start with popular sub-tokens. However, it fails because of two major reasons: 1) the method body representation is not efficient, and 2) the hypothesis (i.e., two methods with similar bodies should have similar names) does not hold sometimes. 

    \item\textbf{RQ4: Where and why do the generation-based approaches work or fail?} \\
    \textbf{Motivation}: Since generation-based approaches are the mainstream method for identifying inconsistent method names, it is worth dedicating more effort to analyzing their strengths and weaknesses to further improve their performance.\\
    \textbf{Answer:} Generation-based approaches, i.e., $CAN$, $MN_{IRE}$, $Cognac$, and $GTNM$, work better on methods with simple bodies and short names. They frequently fail because of the ineffective similarity calculation formula and the immature method name generation techniques. 
    
\end{itemize}

This paper makes the following contributions:
\begin{itemize}
    \item A new, clean benchmark, thoroughly inspected manually, that is large and reflective of real-world scenarios for identifying inconsistent method names. 
    \item An extensive empirical study on the representative DL-based approaches for automated identification of inconsistent method names under different empirical settings.
    \item Key insights that serve as take-away messages, which could be valuable for the future development of advanced approaches.
    
\end{itemize}

The rest of this paper is structured as follows. We review related work in Section~\ref{sec:RelatedWork} and introduce the empirical settings, new dataset construction, and hyperparameter tuning in Section~\ref{sec:ExperimentalSetup}. Sections~\ref{sec:RQ1}-\ref{sec:RQ4} present the methods and result analysis. Section~\ref{sec:Discussion} first discusses threats to validity and limitations. Then a complementarity analysis is presented. Lastly, some insights are also provided for subsequent research in this section. 
Section~\ref{sec:Conclusion} concludes the paper.

\section{Related Work}\label{sec:RelatedWork}
Inconsistent names tend to lead to misunderstandings among developers~\citep{takang1996effects,DBLP:conf/ppig/LiblitBS06,DBLP:journals/ese/ArnaoudovaPA16,hofmeister2017shorter,DBLP:conf/wcre/Arnaoudova10} and may result in incorrect method invocation and software defects~\citep{butler2009relating,amann2018systematic,abebe2011effect,abebe2012can,DBLP:conf/icsm/AghajaniNBL18}. To identify inconsistent method names, a few automatic approaches have been proposed. The task of the identification of inconsistent method names is to identify the method names that do not fully describe the functionality and semantics of their method bodies. The mainstream methods include information retrieval-based approaches, i.e., retrieving from a large code corpus, and generation-based approaches, i.e., generating a method name first, and then validating whether it is consistent with method bodies. We will elaborate on the mainstream approaches in the below subsections. 
There are also other approaches, such as data mining-based approach~\citep{host2009debugging} or classifiers based on DNNs~\citep{NameChecker}. However, since they are not mainstream methods for identifying inconsistent method names, we did not include them in the evaluation in this paper. 

\subsection{Empirical Studies on Identification of Inconsistent Method Names}
Notably, some empirical studies that focus on the topic of inconsistent method names are also presented~\citep{Minehisa2021,Kim2023}. Minehisa et al.~\citep{Minehisa2021} presented a comparative study of the vectorization approaches used in inconsistency detection. This work compares the computational cost and the performance of four different vectorization approaches, proving that Sent2Vec is the best approach which can build a vectorization model 14 times faster than CNN without sacrificing the performance of detecting inconsistent method names. This work only focuses on the vectorization approaches used in the approaches. By contrast, our work extensively evaluated the performance of state-of-the-art approaches in different application scenarios and investigated the reason why they succeed or fail.
Kim et al.~\citep{Kim2023} also presented an empirical study similar to this paper. As far as we know, it is the first work to propose to evaluate the approaches designed to detect inconsistency between method names and bodies from a different perspective based on code review. The authors first constructed a sample benchmark by matching name changes with code reviews. Then they conducted an empirical study on how state-of-the-art approaches perform on this sample benchmark. In addition, they also identified potential biases in the evaluation of SOTA techniques. The major difference between this work and the work presented in this paper is two-fold: First, the rationale of constructing a dataset is different. This work is from the perspective of code review while our work is from a traditional perspective of mining from commit histories combined with manual inspection. Second, the perspective of data analysis is different. This work conducted the data analysis from the perspective of the dataset, indicating that the methods in their benchmark may provide additional information for the correct identification. Besides the analysis based on the dataset, our work further analyzed when and where the evaluated approaches work or fail from different perspectives, e.g., length of method names, initial tokens of method names, complexity of method bodies, and the adopted vectorization techniques.

\subsection{Information Retrieval-based Approaches}
Liu et al.~\citep{Liu2019} proposed a DL-based approach to identify inconsistent method names. To the best of our knowledge, this is the first approach combining deep learning techniques with an information retrieval mechanism for the automated identification of inconsistent method names.

In addition, Liu et al.~\citep{Liu2019} constructed the first benchmark for the task of detecting inconsistent method names, which are extensively used for the evaluations of the subsequent approaches. Consequently, we presented it as a single category.

It converts the method bodies and names in training data into fixed-length vectors by training the models through deep learning techniques, i.e., convolutional neural network (CNN)~\citep{DBLP:journals/nn/MatsuguMMK03}, Word2vec~\citep{DBLP:conf/nips/MikolovSCCD13}, and Paragraph2vec~\citep{DBLP:conf/icml/LeM14}. After the training, the models that can convert method names and bodies into vectors are obtained, and all the converted vectors of method names and bodies in training data constitute the name vector space $vs_{name}$ and body vector space $vs_{body}$, respectively. For a method $<Mbody, Mname>$ to be tested, it first converts $Mbody$ and $Mname$ into vectors using the pre-trained model, then it searches the body vector space $vs_{body}$ for method bodies whose vectors are highly similar to that of $Mbody$ and collects the method names (noted as $MNs_1$) associated with the resulting method bodies. It also searches the name vector space $vs_{name}$ for method names (noted as $MNs_2$) that are both lexically and semantically similar to $Mname$. If $MNs_1$ has no method name that shares the same first sub-token with any method name in $MNs_2$, the test method name will be identified as an inconsistent name.
To evaluate the proposed approach, Liu et al.~\citep{Liu2019} discovered the renaming of methods by mining version control systems. Exactly half of the renaming (randomly selected) was exploited to create positive items (inconsistent method names) by extracting the elder version of the method names. The other half of the renaming was used to create negative items (consistent method names) by extracting the new version of the method names. Their evaluation of the resulting data (noted as $OriginalData$ in the rest of our paper) suggests that their approach is accurate, with a precision of 56.8\% an d a recall of 84.5\%. Besides, Liu et al. conducted a live study on active software projects, and 73 out of 100 pull requests for renaming suggestions were accepted by the developers, which proves the effectiveness of the proposed approach.

\subsection{Generation-based Approaches}~\label{subsec:GenerationApproaches}
The rationale of generation-based approaches is to first generate a method name for a specific method body. Then they calculate the lexical similarity between the generated method name and the original one. Finally, a similarity threshold is adopted to identify whether the original method name is consistent with its method body according to the lexical similarity calculated above. There are three approaches~\citep{Nguyen2020, Li2021, wang2021lightweight} proposed with evaluations on the task of identification of method name consistency. In addition, there are many approaches designed for method name generation without the evaluation of our task. For these approaches, we only consider the latest one~\citep{liu2022learning}, and the pioneering one~\citep{Allamanis2016} as the baselines in this paper.

\emph{$MN_{IRE}$} proposed by ~\citep{Nguyen2020} is another DL-based approach.
To the best of our knowledge, this is the first generation-based approach to automated identification of inconsistent method names. For a method $<Mbody, Mname>$ to be tested, \emph{$MN_{IRE}$} first leverages a deep neural network to generate a method name (noted $N_{g}$) for the given method body $Mbody$. Then it computes the lexical similarity between the generated method name $N_{g}$ and the original one $ Mname$ based on the overlapping rate of sub-tokens. If and only if the similarity
$Sim(N_{g},Mname)$ is smaller than a \emph{threshold} (0.94 in their evaluation), method name $Mname$ will be regarded as inconsistent. One of the key contributions is that \emph{$MN_{IRE}$} leverages additional contexts (e.g., the methods' parameter types, return type, and the enclosing class name) besides the method body to generate the method name.  \emph{$MN_{IRE}$} was evaluated on the dataset $OriginalData$ created by Liu et al.~\citep{Liu2019}. Their evaluation results suggest that \emph{$MN_{IRE}$} outperforms baseline approaches by improving precision from 56.8\% to 62.7\% and recall from 84.5\% to 93.6\%. Besides, in the live study conducted by Nguyen et al., 31 out of 42 pull requests for renaming suggestions were acknowledged by the developers. The evaluation results indicate that \emph{$MN_{IRE}$} works well on the identification of inconsistent method names.

\emph{DeepName} proposed by Li et al.~\citep{Li2021} is another DL-based approach to the automated identification of inconsistent method names. Different from existing approaches, \emph{DeepName} exploits tokens in the caller and callee method of the method under test. Furthermore, they proposed a \emph{Non-copy} mechanism to further improve the generic RNN-based encoder-decoder models. As a result of the additional information and additional mechanism, \emph{DeepName} achieves an F-score of 81.0\% and an accuracy of 75.8\%. 

\emph{Cognac} proposed by Wang et al.~\citep{wang2021lightweight} is the latest DL-based approach for the automated identification of inconsistent method names. Compared with \emph{DeepName}, \emph{Cognac} further leveraged prior knowledge (i.e., probability) of sub-tokens appearing in different contexts besides caller and callee method tokens. In addition, \emph{Cognac} leverages a customized pointer-generator network to incorporate the prior knowledge. The evaluation results suggest \emph{Cognac} achieves an F-score of 80.6\% and an accuracy of 76.6\%. Notably, both evaluations are based on the dataset $OriginalData$ created by Liu et al.~\citep{Liu2019}.

Allamanis et al.~\citep{Allamanis2016} proposed a DL-based approach the generate method names. They leveraged a convolutional attentional network to generate short and descriptive summaries (i.e., method names) for a piece of code snippets. As the first attempt, they generate summary tokens based on the weights of the method body tokens with the help of the attention mechanism and copy mechanism. To the best of our knowledge, this is the first DL-based approach for the method name generation. We take it into our evaluation because it is pioneering and representative.

\emph{GTNM} proposed by Liu et al.~\citep{liu2022learning} is currently the state-of-the-art approach for method name generation. \emph{GTNM} is a transformer-based model that leverages the self-attention mechanism to capture the rich semantic information of method bodies. It initially leveraged the global context of the target method to generate names. The global context includes in-file methods (i.e., other method names in the same file with the target method) and cross-file methods (i.e., the method names in the files imported by the file containing the target method). They also incorporate documentation of methods as a type of context.

We conclude from the preceding analysis that there are some effective approaches to automated identification of inconsistent method names. However, the state-of-the-art DL-based approaches~\citep{Liu2019,Nguyen2020,Li2021,wang2021lightweight} have adopted the same dataset and are only evaluated on the same evaluation setting (i.e., within-project setting and balanced dataset with equivalent inconsistent method names and consistent method names).
Liu et al. take these settings because all the testing methods they extracted have two versions, i.e., the buggy version and the fixed version. Liu et al. label the buggy version methods as inconsistent and the fixed version methods as consistent, which results in a balanced dataset. Nguyen et al.~\citep{Nguyen2020}, Li et al.~\citep{Li2021}, and Wang et al.~\citep{wang2021lightweight} use the same dataset and setting to facilitate the comparison against $IRMCC$. However, it remains unknown whether these DL-based approaches can achieve high accuracy in a widely used setting (cross-project setting) on a more realistic dataset (with a natural ratio of inconsistent method names), which motivates this study.

\section{Experimental Setup}\label{sec:ExperimentalSetup}
\subsection{Evaluated Approaches}
\begin{table*}[]
    \caption{Overview of Existing Approaches.}
    \label{tab:relevantApproaches}
    
    \setlength\tabcolsep{4pt}
    \renewcommand{\arraystretch}{1.2}
    \centering
    \begin{tabular}{cccccc}
    \toprule
    \textbf{Name}                   & \textbf{Publication} & \textbf{Dataset}                               & \textbf{Accuracy}  &\textbf{Classification} &\textbf{Selected}\\
    \midrule

    \makecell[c]{\emph{CAN}}                      & ICML2016                     & \makecell[c]{Self-built dataset \\(11 projects)}                   & --    & Generation &  $\surd$          \\
    \makecell[c]{\emph{IRMCC}}                      & ICSE2019                    & \makecell[c]{Self-built dataset \\(430 projects)}                   & 60.9\%     & IR &  $\surd$          \\
    \makecell[c]{\emph{$MN_{IRE}$}}                         & ICSE2020                     & \makecell[c]{Reuse Liu's Data}                     & 68.9\%         & Generation & $\surd$     \\
    \makecell[c]{\emph{DeepName}}                        & ICSE2021                   & \makecell[c]{Reuse Liu's Data}                     & 75.8\%  &Generation &      \\
    \makecell[c]{\emph{Cognac}}                        & FSE2021                   & \makecell[c]{Reuse Liu's Data}                     & 76.6\%  &Generation &  $\surd$     \\
    \makecell[c]{\emph{GTNM}}                        & ICSE2022                   & \makecell[c]{Reuse Nguyen's Data}                     & --  &Generation &  $\surd$     \\
    \bottomrule
    \end{tabular}
    \end{table*}

An overview of the approaches that are designed to identify inconsistent method names (or are capable of) is presented in Table~\ref{tab:relevantApproaches}. Notably, the first two approaches were not named officially by their authors.
For simplicity's sake, we refer to the first one proposed by Allamanis et al.~\citep{Allamanis2016}, and the second approach proposed by Liu et al.~\citep{Liu2019} as $CAN$, and $IRMCC$, respectively.
Although $CAN$ and $GTNM$ were not evaluated in the original paper, they are still capable of identifying inconsistent method names. We take them into evaluation due to their representativeness which is discussed in Section~\ref{subsec:GenerationApproaches}.
Notably, although the implementation of \emph{DeepName} is publicly available~\citep{Li's}, it cannot run smoothly. Moreover, we contacted the authors and did not get feedback from them. 
Consequently, in this paper, we select $CAN$~\citep{Allamanis2016}, $IRMCC$~\citep{Liu2019}, $MN_{IRE}$~\citep{Nguyen2020}, $Cognac$~\citep{wang2021lightweight}, and $GTNM$~\citep{liu2022learning} for evaluation.

\subsection{Dataset}\label{subsec:dataset_construction}
\subsubsection{Reasons for Building BenMark}
We first explain the reasons why constructing $BenMark$ is necessary. As shown in Table~\ref{tab:relevantApproaches}, the state-of-the-art DL-based approaches~\citep{Liu2019,Nguyen2020,Li2021} are all evaluated on the same dataset, created by~\citep{Liu2019}. In this dataset, called $OriginalData$~\citep{Liu's} for short in this paper, each instance in $OriginalData$ is a triplet, i.e., $<BuggyName, FixedName, MethodBody>$.  Liu et al. take half of the $FixedName$ and $MethodBody$ as consistent methods, and the other half of the $BuggyName$ and $MethodBody$ as inconsistent methods for testing, resulting in a balanced dataset. However, the ratio of inconsistent and consistent methods is extremely imbalanced in real-world scenarios. To investigate how the ratio of inconsistent method names in testing data influences the performance of the selected approaches, we have to construct a new test dataset where the ratios of inconsistent and consistent methods are close to that in real-world applications.  

In addition, as we can see from Table~\ref{tab:originalDataAndBenMark}, the original test dataset, i.e., $OriginalData$, contains only 1,402 inconsistent methods and 1,403 consistent methods, which may lead to unreliable evaluation results according to the findings of Liu et al.~\citep{liu2020deep}. Constructing a substantially larger dataset could be highly valuable for the evaluation. 
As is shown in Table~\ref{tab:originalDataAndBenMark}, $BenMark$ contains 2,443 inconsistent methods and 1,296,743 consistent methods, and the ratio of inconsistent and consistent methods is 1:531, which simultaneously solves the above-mentioned two problems.

For the reasons above, we build $BenMark$ by reusing the subject projects exploited by existing DL-based approaches~\citep{Liu2019,Nguyen2020,Li2021, wang2021lightweight}. There are 430 projects (i.e., GitHub repositories) collected by Liu et al., and they are coming from four well-known communities (namely Apache, Spring, Hibernate, and Google) and have at least 100 commits, making sure that they have been well-maintained. The list of project names and GitHub URLs can be found in~\citep{Liu's}. Detailed construction procedures are shown in Section~\ref{subsubsec:ConstructionOfBenMark}.
    
\begin{table}[]
    \caption{Statistics of Existing Dataset and BenMark.}
    \centering
    
    \label{tab:originalDataAndBenMark}
    \renewcommand{\arraystretch}{1.2}
    \begin{threeparttable}
    \begin{tabular}{@{}ccc@{}}
    \toprule
                     & \textbf{OriginalData}  & \textbf{BenMark} \\ \midrule
    \#Methods(Inc)   & 1,402                      & 2,443                                                           \\
    \#Methods(Con)   & 1,403                     & 1,296,743                                                       \\
    \#Methods(Total) & 2,805                     & 1,299,186     \\
    Ratio & 1:1                   & 1:531                                                \\ \bottomrule
    \end{tabular}
    \begin{tablenotes}    
        \footnotesize 
                    
        \item[a] Inc represents inconsistent methods; 
        \item[b] Con represents consistent methods.       
  
      \end{tablenotes}         
    \end{threeparttable}     
    \end{table}

\subsubsection{Construction of BenMark}
\label{subsubsec:ConstructionOfBenMark}
The construction of $BenMark$ includes two major steps, i.e., automatic identification and manual inspection. We first leveraged the basic logic adopted by Liu et al. to automatically identify inconsistent and consistent method names from projects' commit history. Since the developers could rename a method for various reasons, the above inconsistent method names obtained from renamings in commit history must include noise. To reduce these noise data, we surveyed three developers on how to pick out the genuine inconsistent method names and then performed a manual inspection to maximally reduce the false positives.

\paragraph{\textbf{Automatic Identification}}
The automatic identification of inconsistent and consistent method names are as follows:
\begin{itemize}
    \item \textbf{Data cleaning.}
    Following Liu et al.~\citep{Liu2019}, we exclude \emph{main} methods, empty methods, constructor methods, example methods, and methods with non-alphabetic names. 
    
    \item \textbf{Identifying inconsistent method names.}
    Following Liu et al.~\citep{Liu2019}, we identify inconsistent method names by mining the version control systems. If a method has been renamed (while its body remains unchanged) in a commit and the method remains untouched ever since then, we treat the elder name of the method (i.e., before renaming) as an inconsistent name. Finally, we obtained 4,597 inconsistent methods. Note that Liu et al. filtered out the method names whose first sub-tokens remain the same after renaming. While this filtering is effective in avoiding false positives, it could also miss some true cases. We removed this rule and further validated the data manually. 

    \item \textbf{Identifying consistent method names.}
    We only identify consistent method names in a single snapshot (i.e., the last commit version) of the application. A method in this snapshot is taken as consistent if 1) the method body is not associated with any inconsistent method names identified in the preceding step; 2) both the method body and the method name are untouched during the last $n$ commit versions. For each subject project, $n$ is a constant, i.e., the largest duration between the creation time of a method and its first rename. The value of $n$ for each project can be found in the online appendix~\citep{MCC}. Finally, we obtained 1,296,743 consistent methods.
\end{itemize}

\paragraph{\textbf{Survey}}

Inconsistent method names mined from commit histories could involve various false positives because they can not guarantee the method name changes are associated with the inconsistency between names and bodies~\citep{Wen2020,Wen2022,Kim2023}. To reduce the false positives of the inconsistent method names obtained from the automatic identification, we surveyed developers to investigate how to accurately identify genuine inconsistent method names based on renamings mined from commit histories. 

To reduce the burden on developers, we first conducted an initial inspection of the automatically identified 4,597 inconsistent method names and found three typical cases of false positives: 
    \begin{itemize}
        \item \textbf{Typos correction}, e.g., changing \textbf{``getActoveWebflow''} to \textbf{``getActiveWebflow''};	
        \item \textbf{Format correction}, e.g., changing \textbf{``getTaskId''} to \textbf{``getTaskID''}, or changing \textbf{``reset''} to \textbf{``\_reset''};
        \item \textbf{Add Trailing Number}, e.g., changing \textbf{``getEventFilter''} to \textbf{``getEventFilter0''};	
    \end{itemize}
These false positives are not renamed due to the inconsistency between method names and bodies. After the initial inspection, the number of inconsistent method names is reduced to 4,102.

From the remaining 4,102 inconsistent method names, we randomly sampled 351 inconsistent methods for the survey, with a confidence level of 95\%  and a margin of error of 5\%. 
The participants we invited are three developers working in internet companies, with three, four, and four years of Java development experience, respectively. 
The questions posed to the developers are listed as follows:
\begin{itemize}
    \item First, you should determine whether the given methods were renamed due to inconsistency between method names and bodies based on your development experience.
    \item Second, during the identification, you should try to state the reason behind your judgment and summarize a criterion that can guide the manual identification of inconsistent method names based on renaming instances.
\end{itemize}

Triplets $<BuggyName, FixedName, MethodBody>$ are provided to the developers, and the three developers were given one month to complete the identification and provide a summary of their judgments. Following this, two meetings, each lasting about one hour, were held to discuss discrepancies and refine the judgment criteria.
To assess the reliability of the labeling process, we measure the inter-rater agreement among the three developers by Cohen’s kappa. The kappa value is 0.62, indicating a substantial agreement and highlighting the reliability of the identification.


The results of the survey are presented as follows: For the first question, we found that 237 out of 351 names were renamed due to the inconsistency between names and bodies. However, the remaining 114 names were renamed but not associated with inconsistency between names and bodies, i.e., false positives.  Three typical false positive cases are presented as follows:
    \begin{itemize}
        \item \textbf{Synonyms.} The changed parts (one or multiple sub-tokens) between $BuggyName$ and $FixedName$ are synonyms, e.g., \textbf{``newDocumentBuilder''} is renamed to 
        \textbf{``createDocumentBuilder''} 
        \item \textbf{Abbreviations \& Full Names.} The changed parts between $BuggyName$ and $FixedName$ are a conversion between abbreviations and full names, e.g., \textbf{``readClassAndObject''} is renamed to \textbf{``readClassAndObj''}.
        \item \textbf{Different Word Orders.} The differences between $BuggyName$ and $FixedName$ are only the word orders, e.g., \textbf{``reservationSave''} is renamed to \textbf{``saveReservation''}.
    \end{itemize}

For the second question, we adopted the thematic analysis~\citep{cruzes2011recommended} to characterize how developers express the criteria for the identification of genuine inconsistent method names and what are the possible fine-grained types, according to the following process. (1) We first read and analyzed all the answers to the second question to understand how developers described the reason behind their judgments and identified phrases that expressed the criteria. (2) We carefully reread all the answers and identified phrases to generate initial codes and organize them systematically. (3) After generating the initial codes, we aggregated those with similar meanings and identified an initial theme that represented each cluster. Following this step, all codes were categorized under one of the initial themes, facilitating the identification of any emergent cases. (4) We then reviewed the initial set of themes to identify opportunities for merging similar ones. By clarifying the essence of each theme, we combined similar themes into a new overarching theme or incorporated a theme as a sub-theme. (5) In the last step, we defined the final set of themes. To minimize researcher bias, steps (1) to (4) described above were conducted independently by the first two authors~\citep{runeson2009guidelines}. Following this, a series of meetings was held to resolve conflicts and finalize the assignment of themes (step 5).

Finally, we obtained two requirements to identify a genuine inconsistent method name from a renaming refactoring mined from commit history:
\begin{itemize}
    \item \textbf{Misalignment Between Names and Bodies:} The $BuggyName$ cannot comprehensively reflect the functionality of the $MethodBody$. This is a fundamental requirement for identifying an inconsistent method name. 
    \item \textbf{Semantic Changes After Renaming:}  $BuggyName$ and $FixedName$ should have different semantic meaning. This requirement further ensures developers intend to change the semantics of the method names, validating the genuineness of the inconsistent method names with endorsement from the original developer. 
\end{itemize}

Among 237 cases that satisfy the above two criteria, we found three fine-grained types of genuine inconsistent method names, and we call them \textbf{``Generalize''} type, \textbf{``Narrow''} type, and \textbf{``Change''} type, respectively according to the types of semantic changes. Some typical examples of these three types are presented in Listing~\ref{ThreeTypicalExamples}: 
    \begin{itemize}
        \item \textbf{Generalize (22.4\%)}, one or multiple sub-tokens of $BuggyName$ are removed, resulting in a more generalized identifier, i.e., $FixedName$. For example, \textbf{``getAbsoluteInitTime''} is renamed to \textbf{``getInitTime''}.
        \item \textbf{Narrow (52.7\%)}, one or multiple sub-tokens are added to $BuggyName$, resulting in a more descriptive identifier, i.e., $FixedName$. For example, \textbf{``getAddress''} is renamed to \textbf{``getDestinationAddress''}.
        \item \textbf{Change (24.9\%)}, one or multiple sub-tokens of $BuggyName$ are changed into not related ones (not generalized or narrowed), e.g., \textbf{``getPropertyAssignments''} is renamed to \textbf{``createBuilder''}.  
    \end{itemize}

\paragraph{\textbf{Manual Inspection}}

With the above obtained judgment criterion, the first and second authors manually inspected the remaining 3,751 inconsistent methods, independently. Any discrepancies were discussed through three meetings (each meeting lasted for over two hours). The Cohen's kappa coefficient is 0.75, indicating a substantial agreement between the two authors.
Finally, we obtained 2,443 genuine inconsistent method names based on the renaming refactorings mined from commit histories in total. It is worth noting that there are 1,414 (57.9\%) inconsistent method names belonging to the "Change" type, 192 (7.8\%) inconsistent method names belonging to the "Generalize" type, and 837 (34.3\%) inconsistent method names belonging to the "Narrow" type. The above findings to some extent coincided with the results reported by Peruma et al.~\citep{Peruma2018}. Peruma et al. found that most rename refactorings narrow the meaning of the identifiers where they are applied. Even though we have added several filtering to identify the renaming associated with inconsistency between names and bodies, there are still 34.3\% renamings of the "Narrow" type.

Note that the consistent methods in $BenMark$ are too overwhelming, and manually checking every single method is practically infeasible. To assess the quality of the consistent methods in $BenMark$ and figure out the false positive rate, we randomly sampled methods from 1,296,743 consistent methods with a confidence level of 95\% and a margin of error of 5\%, resulting in a sampled dataset including 383 consistent methods. Two authors manually inspected each sampled method and identified whether the method names were consistent with their corresponding bodies. 
We obtained a perfect agreement (Cohen's kappa coefficient 0.82) between the two authors. For the conflict cases, the two authors discussed until they reached a consensus. The final results suggested that 366/383=95.6\% of the consistent methods are identical to the manually inspected results. That is to say, the false positive rate is only 4.4\%.

\begin{lstlisting}[float,style=JavaStyle,caption={Examples of Three Types of Inconsistent Method Names.},label={ThreeTypicalExamples}]
    // Generalize <getAbsoluteInitTime, getInitTime>
    public final long getInitTime() {
        return initTime;
    }
    
    // Narrow <getAddress, getDestinationAddress>
    public InetSocketAddress getDestinationAddress() {
        return destinationAddress;
    }

    // Change <getPropertyAssignments, createBuilder>

    ClassIntrospectorBuilder createBuilder() {
        return new ClassIntrospectorBuilder(this);
    }

    \end{lstlisting}

\subsubsection{Construction of Training Data}
\label{subsubsec:TrainingDataConstruction}
To maximize the performance of the baselines, we also filtered out the extremely complex methods as Liu et al. did. during the construction of training data. To facilitate the different empirical settings, we further constructed two sets of training data based on $BenMark$:
\begin{itemize}
    \item  \emph{CORPUS\_WP} represents the training data for within-project setting.
    \item  \emph{CORPUS\_CP} represents the training data for cross-project settings.
\end{itemize}
Notably, the evaluated approaches request a large number of projects as a corpus for code retrieval~\citep{Liu2019}, or for the training of method name generation models~\citep{Nguyen2020,Allamanis2016,wang2021lightweight,liu2022learning}. Neither of the evaluated approaches requests any labeling of the methods in the code corpus. To improve the reliability of the results and conclusions, we perform 10-fold cross-validation experiments to reduce the possible bias. The projects of each fold (10\% of all the projects) are used to identify inconsistent and consistent method names for testing, and the projects of the other nine folds of data (90\% of all the projects) are used to construct training data, i.e.,  \emph{CORPUS\_CP}. That is to say, $BenMark$ contains 10 sets of data for training and testing. We extract all the methods of the rest of 90\% of the subject projects at their latest snapshot, i.e., the submitted time of their latest commits, following the state-of-the-art DL-based approaches~\citep{Liu2019,Nguyen2020,Li2021}. 

For \emph{CORPUS\_WP}, we have to consider the timeline of testing and training data since it makes no sense that new data are learned for predicting old data. To this end, we construct the \emph{CORPUS\_WP} by the following steps: 
First, we collected the commits of testing data in the order of the timeline (including inconsistent method names and consistent method names). Second, we located the oldest commit, i.e., the commit that is submitted the earliest among all the commits containing testing data. For convenience, we call this commit $commit_e$.
Finally, we collected training data on the very commit before $commit_e$ to mimic the real scenarios. 
In addition, to avoid any data leakage, we carefully eliminated all the methods included in the testing data from \emph{CORPUS\_WP} to make sure that there is no intersection between \emph{CORPUS\_WP} and the testing data.

\subsubsection{Construction of Testing Data}
\label{subsubsec:TestingDataConstruction}
To facilitate the RQ1 and RQ2, we further construct two new testing datasets (we call $BalancedData$, $NaturalData$) based on $BenMark$. Note that these two datasets are constructed from each fold of $BenMark$, which means that there are 10 $BalancedData$ and 10 $NaturalData$ for all the 10 folds of $BenMark$ respectively. Here are the construction processes: 
\begin{itemize}
    \item $NaturalData$ is constructed by reusing all the testing data of $BenMark$. This dataset is natural because we do not intentionally control the ratio of inconsistent and consistent method names.
    \item To investigate the performance of the evaluated approaches in within-project setting and cross-project setting, we have to construct a balanced dataset containing the same number of inconsistent and consistent method names, i.e., $BalancedData$.  $BalancedData$ is constructed by the following two steps. First, we reused all the inconsistent method names in each fold of $BenMark$ (let the number of inconsistent method names be $N_i$ for each fold $i$), and such samples serve as positive items. Second, we randomly sampled the same number of consistent method names (i.e., $N_i$) from the negative items in each fold of $BenMark$, and such samples serve as the negative items in the resulting testing data. To ensure a fair sample, we made certain that the samples were evenly distributed across projects. Specifically, we sampled consistent method names based on the number of inconsistent method names in each project. In other words, a project with more inconsistent method names should sample a larger number of consistent method names. This dataset is balanced because the number of inconsistent method names is the same as the number of consistent ones. Subsequently, the resulting testing data, i.e., $BalancedData$, is composed of $N_i$ positive items and $N_i$ negative items. Finally, the total number of inconsistent method names in all 10 $BalancedData$ equals the number of inconsistent method names in the $Benchmark$., i.e., $\sum_{i=1}^{10} N_i = 2,443$ (refer Table~\ref{tab:originalDataAndBenMark}). The total number of consistent method names in all 10 $BalancedData$ is also 2,443. 
\end{itemize}

\begin{table}[]
    \caption{Parameters Setting of $IRMCC$.}
    \centering
    
    \label{tab:LIU's}
    \renewcommand{\arraystretch}{1.2}
    \begin{tabular}{ccc||cc}

    \multicolumn{4}{l}{\textbf{Parameters of Paragraph2vec in $IRMCC$}} \\

    \toprule
     & \textbf{Parameter} &\textbf{Value}  & \textbf{Parameter}& \textbf{Value} \\\hline
                                    & Size of vector                       & 300           & Learning rate                        & 0.025                \\
                                & Window size                          & 4
                                                             \\\bottomrule

     \multicolumn{4}{l}{\textbf{Parameters of Word2vec in $IRMCC$}} \\

     \toprule
     & \textbf{Parameter} &\textbf{Value}  & \textbf{Parameter}& \textbf{Value} \\\hline
                              & Size of vector                       & 300       & Learning rate                        & 0.001                           \\
                             & Window size                          & 4
                                                            \\\bottomrule
\multicolumn{2}{l}{\textbf{Parameters of $IRMCC$}} \\

     \toprule
     & \textbf{Parameter} &\textbf{Value}  \\ \hline
                          & k                          & 1
                                                            \\\bottomrule
    \end{tabular}
    \end{table}

\begin{table}[]
    \caption{Parameters Setting of $CAN$.}
    \centering
    
    \label{tab:CAN}
    \renewcommand{\arraystretch}{1.2}
    \begin{tabular}{ccc||cc}

    \multicolumn{4}{l}{\textbf{Parameters of $conv\_attention$ in $CAN$}} \\

    \toprule
     & \textbf{Parameter} &\textbf{Value}  & \textbf{Parameter}& \textbf{Value} \\\hline
     & Embedding size               & 128          & Filters(layer1)                              & 8                \\
     & Window size(layer1)                           & 24      & Filters(layer2)                       & 8        \\
     & Window size(layer2)    & 29 & Dropout               & 0.5   \\
                             & Window size(layer3)                         & 10                                 \\\bottomrule

     \multicolumn{4}{l}{\textbf{Parameters of $copy\_attention$ in $CAN$}} \\

     \toprule
     & \textbf{Parameter} &\textbf{Value}  & \textbf{Parameter}& \textbf{Value} \\\hline
      & Embedding size               & 128          & Filters(layer1)                              & 32                \\
     & Window size(layer1)                           & 18       & Filters(layer2)                       & 16      \\
     & Window size(layer2)    & 19 & Dropout               & 0.4    \\
                             & Window size(layer3)                         & 2                                 \\\bottomrule

    \end{tabular}
    \end{table}

    \begin{table}[]
        \caption{Parameters Setting of $MN_{IRE}$.}
        \centering
        
        \label{tab:MNIRE}
        \renewcommand{\arraystretch}{1.2}
        \begin{tabular}{ccc||cc }
         \toprule
         & \textbf{Parameter} &\textbf{Value}  & \textbf{Parameter}& \textbf{Value} \\\hline
                                  & Learning rate                   & 0.001       & Embedding size                   & 128                          \\
                                  & Batch size                        & 1,024      & Hidden size                          &    256                        \\
                                  & Vocabulary size                          &        50,000 & Threshold & 0.89  \\
                                  \bottomrule
        \end{tabular}
        \end{table}

        \begin{table}[]
        \caption{Parameters Setting of $Cognac$.}
        \centering
        
        \label{tab:Cognac}
        \renewcommand{\arraystretch}{1.2}
        \begin{tabular}{ccc||cc }
         \toprule
         & \textbf{Parameter} &\textbf{Value}  & \textbf{Parameter}& \textbf{Value} \\\hline
                                  & Learning rate                   & 0.15       & Embedding size                   & 150                          \\
                                  & Batch size                        & 120      & Hidden size                          &   400                        \\
                                  & Vocabulary size                          &        85,000 & Threshold & 0.85  \\
                                  \bottomrule
        \end{tabular}
        \end{table}

            \begin{table}[]
        \caption{Parameters Setting of $GTNM$.}
        \centering
        
        \label{tab:GTNM}
        \renewcommand{\arraystretch}{1.2}
        \begin{tabular}{ccc||cc }
         \toprule
         & \textbf{Parameter} &\textbf{Value}  & \textbf{Parameter}& \textbf{Value} \\\hline
                                  & Learning rate                   & 0.0001       & Embedding size                   & 512                          \\
                                  & Batch size                        & 64      & Hidden size                          &    2048                       \\
                                  & Vocabulary size                          &        10,000 & Threshold & 0.85  \\
                                  \bottomrule
        \end{tabular}
        \end{table}
\subsection{Parameter Tuning}
\label{subsec:tuning}
To maximize the potential of the evaluated approaches, we perform hyperparameter tuning for such approaches on a GPU server (OS: Ubuntu 18.04.1; CPU: 32 * Intel(R) Xeon(R) CPU E5-2620 v4 @ 2.10GHz; GPU: 4* TITAN RTX; RAM: 128 GB).

For $IRMCC$, we follow the grid-search tuning approach to tune the parameters of Word2vec and Paragraph2vec, i.e., \emph{Size of vector}, \emph{Learning rate}, and \emph{Window size}. Given that Word2vec and Paragraph2vec are both unsupervised learning approaches, we can only identify the performance by conducting the entire experiments but the whole process is time-consuming. To this end, we empirically selected the next-to-be-tested setting, i.e., selecting the value that yields higher accuracy for a single parameter. In the end, we take the combination which yields the best accuracy value as the final combination of parameters. Parameter $k$ of $IRMCC$  represents the number of the most similar method names and bodies retrieved from the training data. As reported by Liu et al.~\citep{Liu2019}, $k=1$ yields the best performance of $IRMCC$ when identifying inconsistent method names. To this end, we set $k=1$ all through the evaluation in this paper.

For $MN_{IRE}$, we also follow the grid-search tuning approach to tune the parameters with the help of \emph{NNI}~\citep{NNI} which is a widely used toolkit to help developers design and tune machine learning models effectively. We tuned the parameters, i.e., \emph{Learning rate}, \emph{Embedding size}, \emph{Batch size}, \emph{Hidden size}, and \emph{Vocabulary size}. For each of the to-be-tested settings, we train $MN_{IRE}$ with the given setting on nine in ten of the training data and then validate the performance on the validation set (i.e., one in ten of the training data). The combination which yields the highest accuracy is selected as the final parameter.
In addition, $MN_{IRE}$ has an additional hyperparameter named \emph{threshold}. Following Nguyen et al., We tune the \emph{threshold} from 0.85 to 1.0 with a step of 0.01 on the validation set, and finally take the \emph{threshold} which balances the F-scores of identifying consistent and inconsistent method names, i.e., 0.89.

For $CAN$, we tuned the hyperparameters using Bayesian Optimization with Spearmint~\citep{DBLP:conf/nips/SnoekLA12} following Allamanis et al.~\citep{Allamanis2016}. The validation set is one in ten of the training data. As $CAN$ is not originally designed for identifying inconsistent method names, we follow $MN_{IRE}$ and use a specific \emph{threshold} to conduct the identification. The tuning of \emph{threshold} is also conducted on the validation dataset with the same procedure as that of $MN_{IRE}$, and the final value of \emph{threshold} is 0.90.

For $Cognac$ and $GTNM$, we tuned the parameters following the grid-search tuning approach with the help of \emph{NNI}~\citep{NNI}. The included parameters are the same as the ones of $MN_{IRE}$. For \emph{threshold}, we adopted the same tuning strategy, and the final values are both 0.85.

The final parameters of the approaches are presented in Table~\ref{tab:LIU's} - Table~\ref{tab:GTNM}, respectively.

\lstset{
    breaklines,
    numbers=left,
    numberstyle= ,
    basicstyle = \scriptsize,
    keywordstyle= \color{ blue!70},commentstyle=\color{red!50!green!50!blue!50},
    frame=shadowbox,
    rulesepcolor= \color{ red!20!green!20!blue!20},
    xleftmargin=3em,
    xrightmargin=1em,
    aboveskip=1em,
    framexleftmargin=2em,
    captionpos=b
}

\begin{table*}[]
    \caption{The training data and testing data leveraged in three empirical settings.}
    \centering
    
    \label{tab:TrainingAndTestingData}
    \renewcommand{\arraystretch}{1.2}
    \begin{threeparttable}
    \begin{tabular}{@{}ccccc@{}}
    \toprule
            & \multicolumn{2}{c}{\textbf{TestingData}} & \multicolumn{2}{c}{\textbf{TrainingData}} \\
    \midrule
            & \textbf{BalancedData} & \textbf{NaturalData} & \textbf{CORPUS\_WP}  &\textbf{CORPUS\_CP}  \\ \midrule
    Within-Project    & $\bullet$        & -            & $\bullet$          & -   \\
    Cross-Project   & $\bullet$        & -          & -          & $\bullet$\\
    Natural Ratio & -        & $\bullet$          & -  & $\bullet$ \\ \bottomrule
    \end{tabular}
    
    \begin{tablenotes}   
        \footnotesize  
                
        \item[$\bullet$] Data Adopted; 
        \item[-] Data Not Adopted.     
 
      \end{tablenotes}
    \end{threeparttable}   
    \end{table*}
\begin{table*}[]
    \caption{Performance (\%) on Three Different Settings of 10-fold Cross-Validation Experiment of $IRMCC$. }
    \centering
    \label{tab:10fold_19}
    \renewcommand{\arraystretch}{1.2}
    \begin{threeparttable}
    \begin{tabular}{|c|ccc|ccc|ccc|}
    \hline
         & \multicolumn{3}{c|}{Within-Project}                            & \multicolumn{3}{c|}{Cross-Project}                           & \multicolumn{3}{c|}{Natural Ratio}                         \\ \hline
    Fold & \multicolumn{1}{c|}{\makecell[c]{F1\\(Inc)}} & \multicolumn{1}{c|}{\makecell[c]{F1\\(Con)}} & ACC    & \multicolumn{1}{c|}{\makecell[c]{F1\\(Inc)}} & \multicolumn{1}{c|}{\makecell[c]{F1\\(Con)}} & ACC    & \multicolumn{1}{c|}{\makecell[c]{F1\\(Inc)}} & \multicolumn{1}{c|}{\makecell[c]{F1\\(Con)}} & ACC   \\ \hline
    1    & \multicolumn{1}{c|}{54.5}  & \multicolumn{1}{c|}{55.6}  & 55.1 & \multicolumn{1}{c|}{66.8}  & \multicolumn{1}{c|}{52.6}  & 61.0 & \multicolumn{1}{c|}{0.5}   & \multicolumn{1}{c|}{60.1}  & 43.0 
    \\ \hline
    2    & \multicolumn{1}{c|}{44.9}  & \multicolumn{1}{c|}{52.6}  & 49.0 & \multicolumn{1}{c|}{66.7}  & \multicolumn{1}{c|}{50.5}  & 60.2 & \multicolumn{1}{c|}{0.6}   & \multicolumn{1}{c|}{57.7}  & 40.6 \\ \hline
    3    & \multicolumn{1}{c|}{52.9}  & \multicolumn{1}{c|}{52.9}  & 52.9 & \multicolumn{1}{c|}{64.0}  & \multicolumn{1}{c|}{50.4}  & 58.3 & \multicolumn{1}{c|}{0.5}   & \multicolumn{1}{c|}{59.7}  & 42.7 \\ \hline
    4    & \multicolumn{1}{c|}{49.0}  & \multicolumn{1}{c|}{52.3}  & 50.7 & \multicolumn{1}{c|}{63.2}  & \multicolumn{1}{c|}{47.7}  & 56.8 & \multicolumn{1}{c|}{0.4}   & \multicolumn{1}{c|}{56.5}  & 39.5 \\ \hline
    5    & \multicolumn{1}{c|}{52.0}  & \multicolumn{1}{c|}{56.0}  & 54.1 & \multicolumn{1}{c|}{58.3}  & \multicolumn{1}{c|}{46.6}  & 53.2 & \multicolumn{1}{c|}{0.7}   & \multicolumn{1}{c|}{58.1}  & 41.1 \\ \hline
    6    & \multicolumn{1}{c|}{52.8}  & \multicolumn{1}{c|}{49.0}  & 51.0 & \multicolumn{1}{c|}{58.6}  & \multicolumn{1}{c|}{44.0}  & 52.4 & \multicolumn{1}{c|}{0.6}   & \multicolumn{1}{c|}{54.5}  & 37.6 \\ \hline
    7    & \multicolumn{1}{c|}{45.1}  & \multicolumn{1}{c|}{53.8}  & 49.8 & \multicolumn{1}{c|}{60.8}  & \multicolumn{1}{c|}{52.8}  & 57.2 & \multicolumn{1}{c|}{0.5}   & \multicolumn{1}{c|}{64.7}  & 47.9 \\ \hline
    8    & \multicolumn{1}{c|}{51.9}  & \multicolumn{1}{c|}{48.0}  & 50.0 & \multicolumn{1}{c|}{68.9}  & \multicolumn{1}{c|}{52.7}  & 62.4 & \multicolumn{1}{c|}{0.7}   & \multicolumn{1}{c|}{59.0}  &42.0 \\ \hline
    9    & \multicolumn{1}{c|}{47.5}  & \multicolumn{1}{c|}{47.0}  & 47.2 & \multicolumn{1}{c|}{49.5}  & \multicolumn{1}{c|}{39.2}  & 44.8 & \multicolumn{1}{c|}{0.6}   & \multicolumn{1}{c|}{52.2}  & 35.5 \\ \hline
    10   & \multicolumn{1}{c|}{66.4}  & \multicolumn{1}{c|}{62.6}  & 64.6 & \multicolumn{1}{c|}{54.2}  & \multicolumn{1}{c|}{50.3}  & 52.3 & \multicolumn{1}{c|}{0.9}   & \multicolumn{1}{c|}{65.1}  & 48.4 \\ \hline
    Avg.   & \multicolumn{1}{c|}{51.7}  & \multicolumn{1}{c|}{53.0}  & 52.4 & \multicolumn{1}{c|}{61.1}  & \multicolumn{1}{c|}{48.7}  & 55.9 & \multicolumn{1}{c|}{0.6}   & \multicolumn{1}{c|}{58.8}  & 41.8 \\ \hline
    \end{tabular}
    \begin{tablenotes}   
        \footnotesize      
                
        \item[a] Inc represents inconsistent methods; 
        \item[b] Con represents consistent methods.  
      \end{tablenotes}            
    \end{threeparttable}      
    \end{table*}

\begin{table*}[]
\caption{Performance (\%) on Three Different Settings of 10-fold Cross-Validation Experiment of $CAN$. }
    \centering
    \label{tab:10fold_CAN}
    \renewcommand{\arraystretch}{1.2}
    \begin{threeparttable}
\begin{tabular}{|c|ccc|ccc|ccc|}
\hline
     & \multicolumn{3}{c|}{Within-Project}                     & \multicolumn{3}{c|}{Cross-Project}                         & \multicolumn{3}{c|}{Natural Ratio}                      \\ \hline
Fold & \multicolumn{1}{c|}{\makecell[c]{F1\\(Inc)}} & \multicolumn{1}{c|}{\makecell[c]{F1\\(Con)}} & ACC  & \multicolumn{1}{c|}{\makecell[c]{F1\\(Inc)}} & \multicolumn{1}{c|}{\makecell[c]{F1\\(Con)}} & ACC  & \multicolumn{1}{c|}{\makecell[c]{F1\\(Inc)}} & \multicolumn{1}{c|}{\makecell[c]{F1\\(Con)}} & ACC  \\ \hline
1    & \multicolumn{1}{c|}{66.9}    & \multicolumn{1}{c|}{26.4}    & 54.3 & \multicolumn{1}{c|}{71.1}    & \multicolumn{1}{c|}{34.1}    & 59.8& \multicolumn{1}{c|}{0.5}     & \multicolumn{1}{c|}{34.4}    & 21.0  \\ \hline
2    & \multicolumn{1}{c|}{62.0}    & \multicolumn{1}{c|}{31.2}    & 51.0 & \multicolumn{1}{c|}{70.3}    & \multicolumn{1}{c|}{34.0}    & 59.0 & \multicolumn{1}{c|}{0.5}     & \multicolumn{1}{c|}{34.7}    & 21.1 \\ \hline
3    & \multicolumn{1}{c|}{66.9}    & \multicolumn{1}{c|}{26.8}    & 54.4 & \multicolumn{1}{c|}{68.7}    & \multicolumn{1}{c|}{32.7}    & 57.3 & \multicolumn{1}{c|}{0.3}     & \multicolumn{1}{c|}{34.3}    & 20.8  \\ \hline
4    & \multicolumn{1}{c|}{68.5}    & \multicolumn{1}{c|}{27.1}    & 56.0 & \multicolumn{1}{c|}{69.1}    & \multicolumn{1}{c|}{25.3}    & 56.3 & \multicolumn{1}{c|}{0.4}     & \multicolumn{1}{c|}{26.0}    & 15.1 \\ \hline
5    & \multicolumn{1}{c|}{68.3}    & \multicolumn{1}{c|}{30.1}    & 56.4 & \multicolumn{1}{c|}{69.5}    & \multicolumn{1}{c|}{26.9}    & 56.9 & \multicolumn{1}{c|}{0.7}     & \multicolumn{1}{c|}{27.1}    & 15.9 \\ \hline
6    & \multicolumn{1}{c|}{68.1}    & \multicolumn{1}{c|}{27.0}    & 55.6 & \multicolumn{1}{c|}{70.1}    & \multicolumn{1}{c|}{32.6}    & 58.6 & \multicolumn{1}{c|}{0.7}     & \multicolumn{1}{c|}{33.2}    & 20.1 \\ \hline
7    & \multicolumn{1}{c|}{70.0}    & \multicolumn{1}{c|}{37.7}    & 59.5 & \multicolumn{1}{c|}{70.7}    & \multicolumn{1}{c|}{38.9}    & 60.4 & \multicolumn{1}{c|}{0.5}     & \multicolumn{1}{c|}{40.4}    & 25.5  \\ \hline
8    & \multicolumn{1}{c|}{66.8}    & \multicolumn{1}{c|}{29.9}    & 54.9 & \multicolumn{1}{c|}{69.9}    & \multicolumn{1}{c|}{30.0}    & 57.9 & \multicolumn{1}{c|}{0.6}     & \multicolumn{1}{c|}{30.8}    & 18.4 \\ \hline
9    & \multicolumn{1}{c|}{68.2}    & \multicolumn{1}{c|}{24.9}    & 55.3 & \multicolumn{1}{c|}{68.1}    & \multicolumn{1}{c|}{24.6}    & 55.2 & \multicolumn{1}{c|}{0.9}     & \multicolumn{1}{c|}{25.5}    & 14.9 \\ \hline
10   & \multicolumn{1}{c|}{67.7}    & \multicolumn{1}{c|}{38.7}    & 57.7 & \multicolumn{1}{c|}{67.9}    & \multicolumn{1}{c|}{36.0}    & 57.3 & \multicolumn{1}{c|}{1.0}     & \multicolumn{1}{c|}{38.7}    & 24.3 \\ \hline
Avg.   & \multicolumn{1}{c|}{67.3}    & \multicolumn{1}{c|}{30.0}    & 55.5 & \multicolumn{1}{c|}{69.5}    & \multicolumn{1}{c|}{31.5}    & 57.9 & \multicolumn{1}{c|}{0.6}     & \multicolumn{1}{c|}{32.5}    & 19.7 \\ \hline
\end{tabular}
    \begin{tablenotes}   
        \footnotesize
                     
        \item[a] Inc represents inconsistent methods; 
        \item[b] Con represents consistent methods.   
      \end{tablenotes}           
    \end{threeparttable}       
\end{table*}

\begin{table*}[]
    \caption{Performance (\%) on Three Different Settings of 10-fold Cross-Validation Experiment of $MN_{IRE}$. }
    \centering
    \label{tab:10fold_20}
    \renewcommand{\arraystretch}{1.2}
    \begin{threeparttable}
    \begin{tabular}{|c|ccc|ccc|ccc|}
    \hline
         & \multicolumn{3}{c|}{Within-Project}                            & \multicolumn{3}{c|}{Cross-Project}                           & \multicolumn{3}{c|}{Natural Ratio}                       \\ \hline
    Fold & \multicolumn{1}{c|}{\makecell[c]{F1\\(Inc)}} & \multicolumn{1}{c|}{\makecell[c]{F1\\(Con)}} & ACC    & \multicolumn{1}{c|}{\makecell[c]{F1\\(Inc)}} & \multicolumn{1}{c|}{\makecell[c]{F1\\(Con)}} & ACC    & \multicolumn{1}{c|}{\makecell[c]{F1\\(Inc)}} & \multicolumn{1}{c|}{\makecell[c]{F1\\(Con)}} & ACC  \\ \hline
    1    & \multicolumn{1}{c|}{67.4}  & \multicolumn{1}{c|}{37.3}  & 57.1 & \multicolumn{1}{c|}{69.3}  & \multicolumn{1}{c|}{43.2}  & 60.1 & \multicolumn{1}{c|}{0.5}   & \multicolumn{1}{c|}{46.7}  & 30.6  \\ \hline
    2    & \multicolumn{1}{c|}{68.9}  & \multicolumn{1}{c|}{42.2}  & 59.6 & \multicolumn{1}{c|}{71.6}  & \multicolumn{1}{c|}{41.9}  & 61.8 & \multicolumn{1}{c|}{0.6}   & \multicolumn{1}{c|}{43.2}  & 27.7 \\ \hline
    3    & \multicolumn{1}{c|}{66.1}  & \multicolumn{1}{c|}{45.0}  &58.1 & \multicolumn{1}{c|}{69.1}  & \multicolumn{1}{c|}{45.0}  & 60.4 & \multicolumn{1}{c|}{0.5}   & \multicolumn{1}{c|}{48.9}  & 32.5 \\ \hline
    4    & \multicolumn{1}{c|}{68.0}  & \multicolumn{1}{c|}{40.2}  & 58.3 & \multicolumn{1}{c|}{71.1}  & \multicolumn{1}{c|}{43.9}  & 61.8 & \multicolumn{1}{c|}{0.5}   & \multicolumn{1}{c|}{44.4}  & 28.6 \\ \hline
    5    & \multicolumn{1}{c|}{68.4}  & \multicolumn{1}{c|}{40.8}  & 58.8 & \multicolumn{1}{c|}{69.1}  & \multicolumn{1}{c|}{39.0}  & 58.9 & \multicolumn{1}{c|}{0.7}   & \multicolumn{1}{c|}{41.6}  & 26.5 \\ \hline
    6    & \multicolumn{1}{c|}{67.3}  & \multicolumn{1}{c|}{42.5}  & 58.3 & \multicolumn{1}{c|}{69.5}  & \multicolumn{1}{c|}{40.2}  & 59.6 & \multicolumn{1}{c|}{0.7}   & \multicolumn{1}{c|}{42.8}  & 27.4 \\ \hline
    7    & \multicolumn{1}{c|}{63.8}  & \multicolumn{1}{c|}{45.7}  & 56.5 & \multicolumn{1}{c|}{70.9}  & \multicolumn{1}{c|}{49.4}  & 63.0 & \multicolumn{1}{c|}{0.5}   & \multicolumn{1}{c|}{53.1}  & 36.3 \\ \hline
    8    & \multicolumn{1}{c|}{69.1}  & \multicolumn{1}{c|}{42.1}  & 59.7 & \multicolumn{1}{c|}{72.5}  & \multicolumn{1}{c|}{44.9}  & 63.3 & \multicolumn{1}{c|}{0.7}   & \multicolumn{1}{c|}{46.1}  & 30.1 \\ \hline
    9    & \multicolumn{1}{c|}{58.7}  & \multicolumn{1}{c|}{32.4}  & 48.7 & \multicolumn{1}{c|}{65.6}  & \multicolumn{1}{c|}{35.7}  & 55.2 & \multicolumn{1}{c|}{0.9}   & \multicolumn{1}{c|}{39.8}  & 25.1 \\ \hline
    10   & \multicolumn{1}{c|}{62.7}  & \multicolumn{1}{c|}{36.8}  & 53.1 & \multicolumn{1}{c|}{70.6}  & \multicolumn{1}{c|}{40.2}  & 60.6 & \multicolumn{1}{c|}{1.1}   & \multicolumn{1}{c|}{41.9}  & 26.8 \\ \hline
    Avg.   & \multicolumn{1}{c|}{66.0}  & \multicolumn{1}{c|}{40.5}  & 56.8& \multicolumn{1}{c|}{69.9}  & \multicolumn{1}{c|}{42.3}  & 60.5 & \multicolumn{1}{c|}{0.7}   & \multicolumn{1}{c|}{44.8}  & 29.2 \\ \hline
    \end{tabular}
    \begin{tablenotes}   
        \footnotesize  
                
        \item[a] Inc represents inconsistent methods; 
        \item[b] Con represents consistent methods.     
 
      \end{tablenotes}       
    \end{threeparttable}      
    \end{table*}

\begin{table*}[]
    \caption{Performance (\%) on Three Different Settings of 10-fold Cross-Validation Experiment of $Cognac$. }
    \centering
    \label{tab:10fold_Cognac}
    \renewcommand{\arraystretch}{1.2}
    \begin{threeparttable}
    \begin{tabular}{|c|ccc|ccc|ccc|}
    \hline
         & \multicolumn{3}{c|}{Within-Project}                            & \multicolumn{3}{c|}{Cross-Project}                           & \multicolumn{3}{c|}{Natural Ratio}                       \\ \hline
    Fold & \multicolumn{1}{c|}{\makecell[c]{F1\\(Inc)}} & \multicolumn{1}{c|}{\makecell[c]{F1\\(Con)}} & ACC    & \multicolumn{1}{c|}{\makecell[c]{F1\\(Inc)}} & \multicolumn{1}{c|}{\makecell[c]{F1\\(Con)}} & ACC    & \multicolumn{1}{c|}{\makecell[c]{F1\\(Inc)}} & \multicolumn{1}{c|}{\makecell[c]{F1\\(Con)}} & ACC  \\ \hline
    1    & \multicolumn{1}{c|}{66.3}  & \multicolumn{1}{c|}{39.6}  & 56.8 & \multicolumn{1}{c|}{70.6}  & \multicolumn{1}{c|}{34.8}  & 59.5 & \multicolumn{1}{c|}{0.7}   & \multicolumn{1}{c|}{33.6}  & 20.4  \\ \hline
    2    & \multicolumn{1}{c|}{68.7}  & \multicolumn{1}{c|}{35.4}  & 57.8 & \multicolumn{1}{c|}{67.0}  & \multicolumn{1}{c|}{32.0}  & 55.6 & \multicolumn{1}{c|}{0.6}   & \multicolumn{1}{c|}{32.3}  & 19.4 \\ \hline
    3    & \multicolumn{1}{c|}{70.3}  & \multicolumn{1}{c|}{37.0}  & 59.6 & \multicolumn{1}{c|}{68.0}  & \multicolumn{1}{c|}{34.7}  & 57.0 & \multicolumn{1}{c|}{0.6}   & \multicolumn{1}{c|}{28.9}  & 17.1 \\ \hline
    4    & \multicolumn{1}{c|}{69.7}  & \multicolumn{1}{c|}{39.0}  & 59.5 & \multicolumn{1}{c|}{65.4}  & \multicolumn{1}{c|}{22.1}  & 52.0 & \multicolumn{1}{c|}{0.6}   & \multicolumn{1}{c|}{29.3}  & 17.4 \\ \hline
    5    & \multicolumn{1}{c|}{64.7}  & \multicolumn{1}{c|}{22.8}  & 51.5 & \multicolumn{1}{c|}{67.0}  & \multicolumn{1}{c|}{18.3}  & 53.0 & \multicolumn{1}{c|}{0.9}   & \multicolumn{1}{c|}{28.8}  & 17.1 \\ \hline
    6    & \multicolumn{1}{c|}{70.7}  & \multicolumn{1}{c|}{29.0}  & 58.5 & \multicolumn{1}{c|}{67.6}  & \multicolumn{1}{c|}{29.9}  & 55.7 & \multicolumn{1}{c|}{0.3}   & \multicolumn{1}{c|}{30.9}  & 18.4 \\ \hline
    7    & \multicolumn{1}{c|}{63.2}  & \multicolumn{1}{c|}{25.0}  & 50.6 & \multicolumn{1}{c|}{69.2}  & \multicolumn{1}{c|}{32.1}  & 57.6 & \multicolumn{1}{c|}{0.5}   & \multicolumn{1}{c|}{37.4}  & 23.2 \\ \hline
    8    & \multicolumn{1}{c|}{64.9}  & \multicolumn{1}{c|}{12.9}  & 50.0 & \multicolumn{1}{c|}{64.3}  & \multicolumn{1}{c|}{23.3}  & 51.2 & \multicolumn{1}{c|}{0.7}   & \multicolumn{1}{c|}{30.7}  & 18.4 \\ \hline
    9    & \multicolumn{1}{c|}{66.7}  & \multicolumn{1}{c|}{25.7}  & 54.0 & \multicolumn{1}{c|}{68.2}  & \multicolumn{1}{c|}{27.5}  & 55.8 & \multicolumn{1}{c|}{0.8}   & \multicolumn{1}{c|}{26.6}  & 15.7 \\ \hline
    10   & \multicolumn{1}{c|}{68.6}  & \multicolumn{1}{c|}{19.5}  & 54.8 & \multicolumn{1}{c|}{70.4}  & \multicolumn{1}{c|}{35.3}  & 59.4 & \multicolumn{1}{c|}{3.8}   & \multicolumn{1}{c|}{36.3}  & 23.4 \\ \hline
    Avg.   & \multicolumn{1}{c|}{67.4}  & \multicolumn{1}{c|}{28.6}  & 55.3 & \multicolumn{1}{c|}{67.8}  & \multicolumn{1}{c|}{29.0}  & 55.7 & \multicolumn{1}{c|}{1.0}   & \multicolumn{1}{c|}{31.5}  & 19.0 \\ \hline
    \end{tabular}
    \begin{tablenotes}   
        \footnotesize  
                
        \item[a] Inc represents inconsistent methods; 
        \item[b] Con represents consistent methods.     
 
      \end{tablenotes}       
    \end{threeparttable}      
    \end{table*}

\begin{table*}[]
    \caption{Performance (\%) on Three Different Settings of 10-fold Cross-Validation Experiment of $GTNM$. }
    \centering
    \label{tab:10fold_GTNM}
    \renewcommand{\arraystretch}{1.2}
    \begin{threeparttable}
    \begin{tabular}{|c|ccc|ccc|ccc|}
    \hline
         & \multicolumn{3}{c|}{Within-Project}                            & \multicolumn{3}{c|}{Cross-Project}                           & \multicolumn{3}{c|}{Natural Ratio}                       \\ \hline
    Fold & \multicolumn{1}{c|}{\makecell[c]{F1\\(Inc)}} & \multicolumn{1}{c|}{\makecell[c]{F1\\(Con)}} & ACC    & \multicolumn{1}{c|}{\makecell[c]{F1\\(Inc)}} & \multicolumn{1}{c|}{\makecell[c]{F1\\(Con)}} & ACC    & \multicolumn{1}{c|}{\makecell[c]{F1\\(Inc)}} & \multicolumn{1}{c|}{\makecell[c]{F1\\(Con)}} & ACC  \\ \hline
    1    & \multicolumn{1}{c|}{55.9}  & \multicolumn{1}{c|}{35.6}  & 47.7 & \multicolumn{1}{c|}{60.3}  & \multicolumn{1}{c|}{53.6}  & 57.2 & \multicolumn{1}{c|}{0.6}   & \multicolumn{1}{c|}{54.9}  & 38.0  \\ \hline
    2    & \multicolumn{1}{c|}{65.9}  & \multicolumn{1}{c|}{38.1}  & 56.0 & \multicolumn{1}{c|}{64.6}  & \multicolumn{1}{c|}{42.9}  & 56.3 & \multicolumn{1}{c|}{0.4}   & \multicolumn{1}{c|}{48.8}  & 32.3 \\ \hline
    3    & \multicolumn{1}{c|}{66.9}  & \multicolumn{1}{c|}{41.9}  & 57.8 & \multicolumn{1}{c|}{67.7}  & \multicolumn{1}{c|}{45.3}  & 59.4 & \multicolumn{1}{c|}{0.4}   & \multicolumn{1}{c|}{51.6}  & 34.8 \\ \hline
    4    & \multicolumn{1}{c|}{67.4}  & \multicolumn{1}{c|}{34.5}  & 56.5 & \multicolumn{1}{c|}{65.3}  & \multicolumn{1}{c|}{36.8}  & 55.2 & \multicolumn{1}{c|}{0.4}   & \multicolumn{1}{c|}{48.2}  & 31.9 \\ \hline
    5    & \multicolumn{1}{c|}{68.1}  & \multicolumn{1}{c|}{31.8}  & 56.5 & \multicolumn{1}{c|}{68.0}  & \multicolumn{1}{c|}{44.3}  & 59.4 & \multicolumn{1}{c|}{0.7}   & \multicolumn{1}{c|}{47.2}  & 31.0 \\ \hline
    6    & \multicolumn{1}{c|}{71.5}  & \multicolumn{1}{c|}{48.4}  & 63.3 & \multicolumn{1}{c|}{68.4}  & \multicolumn{1}{c|}{51.5}  & 61.7 & \multicolumn{1}{c|}{0.2}   & \multicolumn{1}{c|}{51.9}  & 35.1 \\ \hline
    7    & \multicolumn{1}{c|}{61.2}  & \multicolumn{1}{c|}{36.9}  & 52.0 & \multicolumn{1}{c|}{67.1}  & \multicolumn{1}{c|}{50.7}  & 60.5 & \multicolumn{1}{c|}{0.4}   & \multicolumn{1}{c|}{60.2}  & 43.2 \\ \hline
    8    & \multicolumn{1}{c|}{63.2}  & \multicolumn{1}{c|}{35.5}  & 53.1 & \multicolumn{1}{c|}{65.6}  & \multicolumn{1}{c|}{41.3}  & 56.6 & \multicolumn{1}{c|}{0.7}   & \multicolumn{1}{c|}{48.3}  & 32.0 \\ \hline
    9    & \multicolumn{1}{c|}{64.9}  & \multicolumn{1}{c|}{33.8}  & 54.2 & \multicolumn{1}{c|}{62.7}  & \multicolumn{1}{c|}{31.1}  & 51.6 & \multicolumn{1}{c|}{0.7}   & \multicolumn{1}{c|}{42.3}  & 27.0 \\ \hline
    10   & \multicolumn{1}{c|}{72.7}  & \multicolumn{1}{c|}{62.9}  & 68.5 & \multicolumn{1}{c|}{74.2}  & \multicolumn{1}{c|}{52.6}  & 66.6 & \multicolumn{1}{c|}{3.1}   & \multicolumn{1}{c|}{53.3}  & 37.0 \\ \hline
    Avg.   & \multicolumn{1}{c|}{65.8}  & \multicolumn{1}{c|}{39.9}  & 56.6& \multicolumn{1}{c|}{66.4}  & \multicolumn{1}{c|}{45.0}  & 58.4 & \multicolumn{1}{c|}{0.8}   & \multicolumn{1}{c|}{50.7}  & 34.2 \\ \hline
    \end{tabular}
    \begin{tablenotes}   
        \footnotesize  
                
        \item[a] Inc represents inconsistent methods; 
        \item[b] Con represents consistent methods.     
 
      \end{tablenotes}       
    \end{threeparttable}      
    \end{table*}

    \section{RQ1: Cross-Project VS. Within-Project} \label{sec:RQ1}
    To answer this research question, we evaluate the selected DL-based approaches with cross-project and within-project settings, independently. Both within-project settings and cross-project settings are valuable in different application scenes. For within-project settings, instead of ignoring all data from the target project, developers can put the confirmed consistent data into training data to train a new model, and then further conduct the prediction on the remaining data. However, this requires repetitious training of the models, which is often time-consuming and resource-consuming. For cross-project settings, users can train the model once and use this model to predict any new test data. Therefore, they do not have to train models on each new project they encounter, which significantly facilitates usage. By comparing the performance of the same DL-based approaches under two empirical settings, we can reveal whether the evaluated approaches can also achieve promising performance in cross-project settings. The adopted training data and testing data are presented in Table~\ref{tab:TrainingAndTestingData}.

    \subsection{Process}
    
    \subsubsection{Cross-Project Evaluation}\label{subsub:Cross-project}
    The testing data is $BalancedData$ (see Section~\ref{subsubsec:TestingDataConstruction}).
   The training dataset (i.e., \emph{CORPUS\_CP}, see Section~\ref{subsubsec:TrainingDataConstruction}) is constructed over the rest of 90\% of all the projects, i.e., the other 9 fold of data.
    This dataset is leveraged to train the method name generation model of $MN_{IRE}$, $CAN$, $Cognac$, and $GTNM$, and it also serves as the code repository requested by $IRMCC$.

    The evaluation is cross-project because the methods in the testing dataset (i.e., $BalancedData$) and the methods in the training data (i.e., \emph{CORPUS\_CP}) are extracted from different projects.
    
    \subsubsection{Within-Project Evaluation}
    The testing dataset in within-project settings is the same as that in cross-project settings, i.e., $BalancedData$.
    The training dataset is \emph{CORPUS\_WP} (see Section~\ref{subsubsec:TrainingDataConstruction})
    The evaluation is within-project settings because the methods in the testing dataset (i.e., $BalancedData$) and the methods in the training dataset (i.e., \emph{CORPUS\_WP}) may come from the same project.
    Notably, cross-project evaluation and within-project evaluation use the same subject projects selected by Liu et al.~\citep{Liu2019}. The major difference is that cross-project evaluation partitions projects into testing and training projects that are exploited to create testing data and training data respectively. In contrast, within-project evaluation selects some methods as testing data, and others are taken as training data regardless of where they come from.

            \begin{table}[htb]
                \caption{Within-Project \emph{vs.} Cross-Project ($IRMCC$).}
                \centering
                
                \renewcommand{\arraystretch}{1.2}
                \label{tab:Comparison19_1_1}
                \begin{tabular}{@{}cccc@{}}
                \toprule
                &{\textbf{Metrics}} & \textbf{Within-Project} & \textbf{Cross-Project} \\ \midrule
                \multirow{3}{*}{\textbf{Inconsistent}} & Precision       & 52.3\%     & 54.4\%         \\
                                                        & Recall         & 51.3\%     & 70.0\%         \\
                                                        & F-score        & 51.7\%     & 61.1\%         \\ \midrule
                \multirow{3}{*}{\textbf{Consistent}} & Precision         & 52.6\%     & 59.0\%         \\
                                                    & Recall             & 53.6\%     & 41.7\%        \\
                                                    & F-score            & 53.0\%     & 48.7\%         \\ \midrule
                \multicolumn{2}{c}{Accuracy}                             & 52.4\%     & 55.9\%         \\ \bottomrule
                \end{tabular}
            \end{table}
    
    \begin{table}[htb]
        \caption{Within-Project \emph{vs.} Cross-Project ($CAN$). }
        \centering
        
        \renewcommand{\arraystretch}{1.2}
        \label{tab:Comparison_CAN_1_1}
        \begin{tabular}{@{}cccc@{}}
        \toprule
        &\textbf{Metrics} & \textbf{Within-Project} & \textbf{Cross-Project} \\ \midrule
        \multirow{3}{*}{\textbf{Inconsistent}} & Precision     & 53.2\%        & 54.5\%\\
                                                & Recall        & 91.8\%        & 96.2\%\\
                                                & F-score        & 67.3\%       & 69.5\%\\ \midrule
        \multirow{3}{*}{\textbf{Consistent}}    & Precision     & 71.6\%        & 84.6\%\\
                                                & Recall        & 19.2\%        & 19.5\%\\
                                                & F-score       & 30.0\%        & 31.5\%\\ \midrule
        \multicolumn{2}{c}{Accuracy}                            & 55.5\%        & 57.9\%
        \\ \bottomrule
        \end{tabular}
    \end{table}

    \begin{table}[htb]
        \caption{Within-Project \emph{vs.} Cross-Project ($MN_{IRE}$).}
        \centering
        
        \renewcommand{\arraystretch}{1.2}
        \label{tab:Comparison20_1_1}
        \begin{tabular}{@{}cccc@{}}
        \toprule
        &\textbf{Metrics} & \textbf{Within-Project} & \textbf{Cross-Project} \\ \midrule
        \multirow{3}{*}{\textbf{Inconsistent}} & Precision       & 54.3\%        & 56.4\%\\
                                                & Recall         & 84.2\%        & 91.9\%\\
                                                & F-score        & 66.0\%        & 69.9\%\\ \midrule
        \multirow{3}{*}{\textbf{Consistent}}    & Precision      & 66.1\%        & 78.7\%\\
                                                & Recall         & 29.4\%        & 29.1\%\\
                                                & F-score        & 40.5\%        & 42.3\%\\ \midrule
        \multicolumn{2}{c}{Accuracy}                             & 56.8\%        & 60.5\%
        \\ \bottomrule
        \end{tabular}
    \end{table}

       \begin{table}[htb]
        \caption{Within-Project \emph{vs.} Cross-Project ($Cognac$).}
        \centering
        
        \renewcommand{\arraystretch}{1.2}
        \label{tab:Comparison_Cognac_1_1}
        \begin{tabular}{@{}cccc@{}}
        \toprule
        &\textbf{Metrics} & \textbf{Within-Project} & \textbf{Cross-Project} \\ \midrule
        \multirow{3}{*}{\textbf{Inconsistent}} & Precision       & 53.1\%        & 53.3\%\\
                                                & Recall         & 92.3\%        & 93.1\%\\
                                                & F-score        & 67.4\%        & 67.8\%\\ \midrule
        \multirow{3}{*}{\textbf{Consistent}}    & Precision      & 71.8\%        & 72.8\%\\
                                                & Recall         & 18.3\%        & 18.2\%\\
                                                & F-score        & 28.6\%        & 29.0\%\\ \midrule
        \multicolumn{2}{c}{Accuracy}                             & 55.3\%        & 55.7\%
        \\ \bottomrule
        \end{tabular}
    \end{table}

    \begin{table}[htb]
        \caption{Within-Project \emph{vs.} Cross-Project ($GTNM$).}
        \centering
        
        \renewcommand{\arraystretch}{1.2}
        \label{tab:Comparison_GTNM_1_1}
        \begin{tabular}{@{}cccc@{}}
        \toprule
        &\textbf{Metrics} & \textbf{Within-Project} & \textbf{Cross-Project} \\ \midrule
        \multirow{3}{*}{\textbf{Inconsistent}} & Precision       & 54.3\%        & 55.1\%\\
                                                & Recall         & 83.6\%        & 83.7\%\\
                                                & F-score        & 65.8\%        & 66.4\%\\ \midrule
        \multirow{3}{*}{\textbf{Consistent}}    & Precision      & 65.1\%        & 68.0\%\\
                                                & Recall         & 29.5\%        & 33.8\%\\
                                                & F-score        & 39.9\%        & 45.0\%\\ \midrule
        \multicolumn{2}{c}{Accuracy}                             & 56.6\%        & 58.4\%
        \\ \bottomrule
        \end{tabular}
    \end{table}

    \subsection{Results}
    Following existing evaluations~\citep{Liu2019}, we leverage the precision, recall, and F-score of identifying positive and negative items as well as the overall accuracy for all testing items. Due to limited space, we only report the F-score of identifying positive and negative items and overall accuracy for the 10-fold cross-validation experiments. The complete performance is available in our public repository~\citep{MCC}. The performance of each fold of data with within-project setting and cross-project setting is presented in Table~\ref{tab:10fold_19} - Table~\ref{tab:10fold_GTNM}. The average evaluation results of 10-fold cross-validation experiments are presented in Table~\ref{tab:Comparison19_1_1} -  Table~\ref{tab:Comparison_GTNM_1_1}, and from these tables we make the following observations:

    \begin{itemize}
      \item First, switching from the within-project setting to the cross-project setting slightly improves the performance of $IRMCC$. The overall accuracy is improved by 3.5 percentage points. We also notice that the performance is reduced for consistent method names (negative items) while the performance is improved for inconsistent ones (positive items). The reduction in F-score is 4.3\%=53.0\%-48.7\% for negative items, and the F-score improves 9.4\%=61.1\%-51.7\% for positive items.
      \item Second, the performance of generation-based approaches, i.e., $CAN$, $MN_{IRE}$, $Cognac$, and $GTNM$, also slightly improved when switching from the within-project setting to the cross-project setting. The overall accuracy is improved by 2.4, 3.7, 0.4, and 1.8 percentage points for the generation-based approaches, respectively. Compared to the performance of $IRMCC$, the F-score of identifying consistent and inconsistent method names both increases.
      \item The slightly improved performance may be attributed to the larger size of \emph{CORPUS\_CP} compared to \emph{CORPUS\_WP}. With more training data, IR-based approach and generation-based approaches are trained more adequately, thus achieving a better performance. In addition, the reason why the performance of identifying inconsistent and consistent method names changes differently is that:  for generation-based approaches, the ability of name generation impacts the performance of identifying inconsistent and consistent method names consistently since the quality of generated names is essential in the subsequent identification. However, the IR-based approach, i.e., $IRMCC$ has to strike a balance between the accuracy of identifying inconsistent and consistent method names.
    \end{itemize}
    
    \begin{tcolorbox}
        \textit{Summary:}
        The performance of the evaluated approaches is slightly improved when switching from a within-project to a cross-project setting, which indicates that the evaluated approaches can also achieve promising performance in a cross-project setting. 
    \end{tcolorbox}

    \section{RQ2: Artificial Data VS. Natural Data}\label{sec:RQ2}

    \subsection{Process}
    
    To answer RQ2, we should evaluate the selected DL-based approaches multiple times by changing the ratio of inconsistent method names in the testing data. Notably, the evaluation of the DL-based approach is time-consuming. Consequently, we only investigate two special settings of the ratio.
    The first setting is to balance positive and negative items in the testing dataset. Subsequently, the testing data is $BalancedData$, and we have experimented on this dataset (see Section~\ref{sec:RQ1}). Given that the performance of evaluated approaches with within-project setting and cross-project setting do not have significant difference (see Section~\ref{sec:RQ1}), we take the performance of cross-project setting in Section~\ref{sec:RQ1} as the performance of $BalancedData$ for comparison because the cross-project setting is wildly used and time-saving for the potential users. To make sure that there is only one independent variable, i.e., the ratio of inconsistent method names, the evaluation in this research question is also based on a cross-project setting. 
    The second special setting is to simulate the ratio of inconsistent and consistent method names in real scenarios. We construct a corresponding dataset, i.e., $NaturalData$ (see Section~\ref{subsubsec:TestingDataConstruction}). This dataset is natural because the number of negative items (consistent method names) is significantly more than positive ones (inconsistent method names).
    
    Notably, the two settings share the same training data (i.e., \emph{CORPUS\_CP}).
    The only difference (variable) among the two settings is the ratio of positive and negative items in the testing data. As a result, we may reveal the effect of the ratio by analyzing how the evaluated approaches perform in these two settings. The adopted training data and testing data are presented in Table~\ref{tab:TrainingAndTestingData}.

    \subsection{Results}
    The performance of each fold of data in different settings is presented in Table~\ref{tab:10fold_19} - Table~\ref{tab:10fold_GTNM}. Table~\ref{tab:Comparison19_real} - 
 Table~\ref{tab:comparison_GTNM_real} present the average evaluation results, and from these tables, we make the following observations:
    
        \begin{itemize}
        \item First, the precision of the evaluated approaches on inconsistent names is significantly impacted when we change the ratio of inconsistent and consistent method names in testing data. Switching from $BalancedData$ to $NaturalData$ (the ratio of inconsistent method names declines sequentially) results in a dramatic decline in precision. The precision is reduced substantially from 54.4\% to 0.3\% ($IRMCC$), from 54.5\% to 0.3\% ($CAN$), from 56.4\% to 0.3\% ($MN_{IRE}$), from 53.3\% to 0.5\%($Cognac$), and from 55.1\% to 0.4\%($GTNM$). The reason is that the more consistent methods we include, the more false positives we get. As a result, the number of true positives is stable whereas false positives are increased, which results in a reduced precision that equals true positives divided by the sum of true positives and false positives.  
       \item Second, the decline of the ratio of inconsistent method names leads to better precision in identifying consistent method names. The precision is increased from 59.0\% to 99.8\% ($IRMCC$), from 84.6\% to 99.9\% ($CAN$), from 78.7\% to 99.9\% ($MN_{IRE}$), from 72.8\% to 99.9\%($Cognac$), and from 68.0\% to 99.9\%. This is because the number of inconsistent method names is too tiny, and it can barely impact the overwhelming number of consistent method names.  
       \item Third, the recall is all slightly changed. The reason is that the evaluated approaches are trained with the same training data regardless of the difference in testing data.  As a result, the resulting models (approaches) have the same ability to identify inconsistent and consistent methods. For example, from Table~\ref{tab:comparison_GTNM_real}, we observe that $GTNM$ can accurately classify 83.7\% of the inconsistent names in $BalancedData$. Increasing the number of consistent names (in $NaturalData$) would not influence the fact that around 83.7\% of the inconsistent names are classified correctly (which should result in a recall of around 83.7\%). Notably, Because of the random sampling during dataset construction and high randomness of DL-based approach~\citep{DBLP:journals/widm/ScardapaneW17,ZhuangRan}, the recall in $BalancedData$, and $NaturalData$ is not the same.
    \end{itemize}

        \begin{table}
        \caption{$BalancedData$ \emph{vs.} $NaturalData$ ($IRMCC$).}
        \centering
        
        \renewcommand{\arraystretch}{1.2}
        \label{tab:Comparison19_real}
        \begin{tabular}{@{}cccc@{}}
        \toprule
        &{\textbf{Metrics}} & \textbf{BalancedData} & \textbf{NaturalData} \\ \midrule
        \multirow{3}{*}{\textbf{Inconsistent}} & Precision     & 54.4\%         & 0.3\%     \\
                                                & Recall        & 70.0\%         & 70.0\%    \\
                                                & F-score      & 61.1\%         & 0.6\%    \\  \midrule
        \multirow{3}{*}{\textbf{Consistent}}   & Precision      & 59.0\%         & 99.8\%    \\
                                                & Recall      & 41.7\%         & 41.7\%    \\
                                                & F-score     & 48.7\%         & 58.8\%     \\ \midrule
        \multicolumn{2}{c}{Accuracy}                             & 55.9\%         & 41.8\%  
        \\ \bottomrule
        \end{tabular}
        \end{table}
    
        \begin{table}
            \caption{$BalancedData$ \emph{vs.} $NaturalData$ ($CAN$).}
            \centering
            
            \renewcommand{\arraystretch}{1.2}
            \label{tab:comparison_CAN_real}
            \begin{tabular}{@{}cccc@{}}
            \toprule
                                            &{\textbf{Metrics}}  & \textbf{BalancedData} & \textbf{NaturalData} \\ \midrule
                                            \multirow{3}{*}{\textbf{Inconsistent}}
                                            & Precision      & 54.5\%        & 0.3\% \\
                                            & Recall       & 96.2\%        &  96.2\% \\
                                            & F-score         & 69.5\%        & 0.6\% \\ \midrule
        \multirow{3}{*}{\textbf{Consistent}}    & Precision      & 84.6\%        & 99.9\% \\
                                            & Recall        & 19.5\%        & 19.5\% \\
                                            & F-score       & 31.5\%        & 32.5\%\\ \midrule
        \multicolumn{2}{c}{Accuracy}                           & 57.9\%        & 19.7\%  \\\bottomrule
            \end{tabular}
            \end{table}

        \begin{table}
            \caption{$BalancedData$ \emph{vs.} $NaturalData$ ($MN_{IRE}$).}
            \centering
            
            \renewcommand{\arraystretch}{1.2}
            \label{tab:comparison20_real}
            \begin{tabular}{@{}cccc@{}}
            \toprule
                                            &{\textbf{Metrics}}  & \textbf{BalancedData} & \textbf{NaturalData} \\ \midrule
                                            \multirow{3}{*}{\textbf{Inconsistent}}
                                            & Precision       & 56.4\%        & 0.3\%\\
                                            & Recall         & 91.9\%        &  92.0\%\\
                                            & F-score       & 69.9\%        & 0.7\%\\ \midrule
        \multirow{3}{*}{\textbf{Consistent}}    & Precision     & 78.7\%        & 99.9\%\\
                                            & Recall        & 29.1\%        & 29.0\%\\
                                            & F-score       & 42.3\%        & 44.8\%\\ \midrule
        \multicolumn{2}{c}{Accuracy}                          & 60.5\%        & 29.2\% \\\bottomrule
            \end{tabular}
            \end{table}

\begin{table}
            \caption{$BalancedData$ \emph{vs.} $NaturalData$ ($Cognac$).}
            \centering
            
            \renewcommand{\arraystretch}{1.2}
            \label{tab:comparison_Cognac_real}
            \begin{tabular}{@{}cccc@{}}
            \toprule
                                            &{\textbf{Metrics}}  & \textbf{BalancedData} & \textbf{NaturalData} \\ \midrule
                                            \multirow{3}{*}{\textbf{Inconsistent}}
                                            & Precision       & 53.3\%        & 0.5\%\\
                                            & Recall         & 93.1\%        &  93.3\%\\
                                            & F-score       & 67.8\%        & 1.0\%\\ \midrule
        \multirow{3}{*}{\textbf{Consistent}}    & Precision     & 72.8\%        & 99.9\%\\
                                            & Recall        & 18.2\%        & 18.7\%\\
                                            & F-score       & 29.0\%        & 31.5\%\\ \midrule
        \multicolumn{2}{c}{Accuracy}                          & 55.7\%        & 19.0\% \\\bottomrule
            \end{tabular}
            \end{table}

       \begin{table}
            \caption{$BalancedData$ \emph{vs.} $NaturalData$ ($GTNM$).}
            \centering
            
            \renewcommand{\arraystretch}{1.2}
            \label{tab:comparison_GTNM_real}
            \begin{tabular}{@{}cccc@{}}
            \toprule
                                            &{\textbf{Metrics}}  & \textbf{BalancedData} & \textbf{NaturalData} \\ \midrule
                                            \multirow{3}{*}{\textbf{Inconsistent}}
                                            & Precision       & 55.1\%        & 0.4\%\\
                                            & Recall         & 83.7\%        &  83.7\%\\
                                            & F-score       & 66.4\%        & 0.8\%\\ \midrule
        \multirow{3}{*}{\textbf{Consistent}}    & Precision     & 68.0\%        & 99.9\%\\
                                            & Recall        & 33.8\%        & 34.1\%\\
                                            & F-score       & 45.0\%        & 50.7\%\\ \midrule
        \multicolumn{2}{c}{Accuracy}                          & 58.4\%        & 34.2\% \\\bottomrule
            \end{tabular}
            \end{table}

    \begin{tcolorbox}
        \textit{Summary:} The performance of the evaluated approaches is impacted substantially by the ratio of inconsistent and consistent method names. This finding indicates that we should evaluate approaches on the dataset that are constructed according to the real scenarios, and existing DL-based approaches for method name consistency checking may not work accurately in the field.
    \end{tcolorbox}

\section{RQ3: Analysis of IR-based Approaches}\label{sec:RQ3}
In Section~\ref{sec:RQ2}, we learn that $IRMCC$ works in many cases (the accuracy can be 41.8\%). However, it remains unclear where and why it works or fails. Answering these questions could shed light on the direction of developing more advanced DL-based approaches. To this end, in this research question, we investigate in what cases and for what reason the IR-based approach, i.e., $IRMCC$, works or fails. 

To answer this research question, we follow the widely-used mixed-method approach~\citep{creswell2017research} and combine a qualitative analysis of the sampled methods with a quantitative examination of the whole set of testing data. On the qualitative analysis, we first randomly sample 383 methods (resulting in 185 successfully identified cases and 198 failed ones) from all the methods in $NaturalData$, with a confidence level of 95\% and a margin of error of 5\%. We analyze where and why $IRMCC$ works or fails from the perspectives of method name and body, with a qualitative analysis based on the rationale of $IRMCC$. From the method name perspective, we analyze the popularity of the first sub-token of method names since $IRMCC$ considers the first sub-token to conduct consistency checking. From the method body perspective, we analyze the code complexity by measuring lines of code (noted as LOC) and McCabe's Cyclomatic Complexity. At the last, for the qualitative analysis, we analyze the names and bodies of the methods retrieved by $IRMCC$.\\

\subsection{Popularity of Method Name's F-token Matters}\label{subsec:FirstTokenAnalysis}
Given that $IRMCC$ heavily relies on the first sub-token of a method name to conduct the consistency checking, we analyze the first sub-tokens (noted as \emph{F-token}) of the sampled 383 method names to investigate how the \emph{F-token}s affect the performance of $IRMCC$ from the method name perspective.

\begin{table}[]
    \caption{Proportions of the Top Three Popular \emph{F-token}s.}
    \centering
    
    \renewcommand{\arraystretch}{1.2}
    \label{tab:FirstTokenAnalysis}
    \begin{tabular}{@{}ccccc@{}}
    \toprule
     &
      \begin{tabular}[c]{@{}c@{}} \textbf{Success} \\ \textbf{Cases}\end{tabular} &
      \begin{tabular}[c]{@{}c@{}} \textbf{Failure} \\ \textbf{Cases}\end{tabular} &
      \begin{tabular}[c]{@{}c@{}} \textbf{Sampled} \\ \textbf{Cases}\end{tabular} &
      \begin{tabular}[c]{@{}c@{}} \textbf{NaturalData}\end{tabular} \\ \midrule
    get* & 51.9\% & 17.2\% & 33.9\% & 31.9\% \\
    set* & 19.5\% & 12.1\% & 15.7\% & 11.6\% \\
    is*  & 4.9\%  & 4.0\%  & 4.4\%  & 4.9\%  \\ \bottomrule
    \end{tabular}
    \end{table}

We collect the \emph{F-token}s of the sampled methods and sort them by their popularity in successfully identified cases, failed identification cases, and all sampled cases, respectively. For comparison, we also sort the \emph{F-token}s by their popularity in $NaturalData$. Table~\ref{tab:FirstTokenAnalysis} shows the top three most frequent \emph{F-token}s in these four scenarios.
From Table~\ref{tab:FirstTokenAnalysis}, we make the following observations:
\begin{itemize}
    \item First, the top three most popular \emph{F-token}s in success cases, failure cases, all sampled cases, and $NaturalData$ keep the same, i.e., \emph{get}, \emph{set}, and \emph{is}.
    \item Second, in the sampled cases identified successfully by $IRMCC$, the ratio of methods starting with the top three \emph{F-token}s is significantly higher than their ratios in the sampled failure cases (34.7\% for \emph{get}, 7.4\% for \emph{set}, and 0.9\% for \emph{is}), and even all sampled cases (18.0\% for \emph{get}, 3.8\% for \emph{set}, and 0.5\% for \emph{is}), which indicates a higher success rate.
    \item Third, with the decrease of the popularity of \emph{F-token}s (i.e., 31.9\%, 11.6\%, and 4.9\%) in $NaturalData$, the difference of ratios between the sampled success cases and sampled failure cases also decreases, i.e., 34.7\%=51.9\%-17.2\%, 7.4\%=19.5\%-12.1\%, and 0.9\%=4.9\%-4.0\%.
\end{itemize}

Through the observations above, we assume that there might be a correlation between the popularity of \emph{F-token}s and their rates of being identified successfully (noted as \emph{success rate}). Therefore, we calculate the \emph{success rate} of all the method names, and we found that the \emph{success rate} of the methods that start with the top three \emph{F-token}s are also the highest ones, i.e., 73.8\%=96/130 for \emph{get*}, 60.0\%=36/60 for \emph{set*}, and 52.9\%=9/17 for \emph{is*}. In contrast, the average success rate (i.e., Accuracy) of the whole dataset is only 48.3\% (see Table~\ref{tab:10fold_19}).

\begin{figure}
    \centering
    \includegraphics[width=0.8\textwidth ,trim={30 30 30 30}, clip]{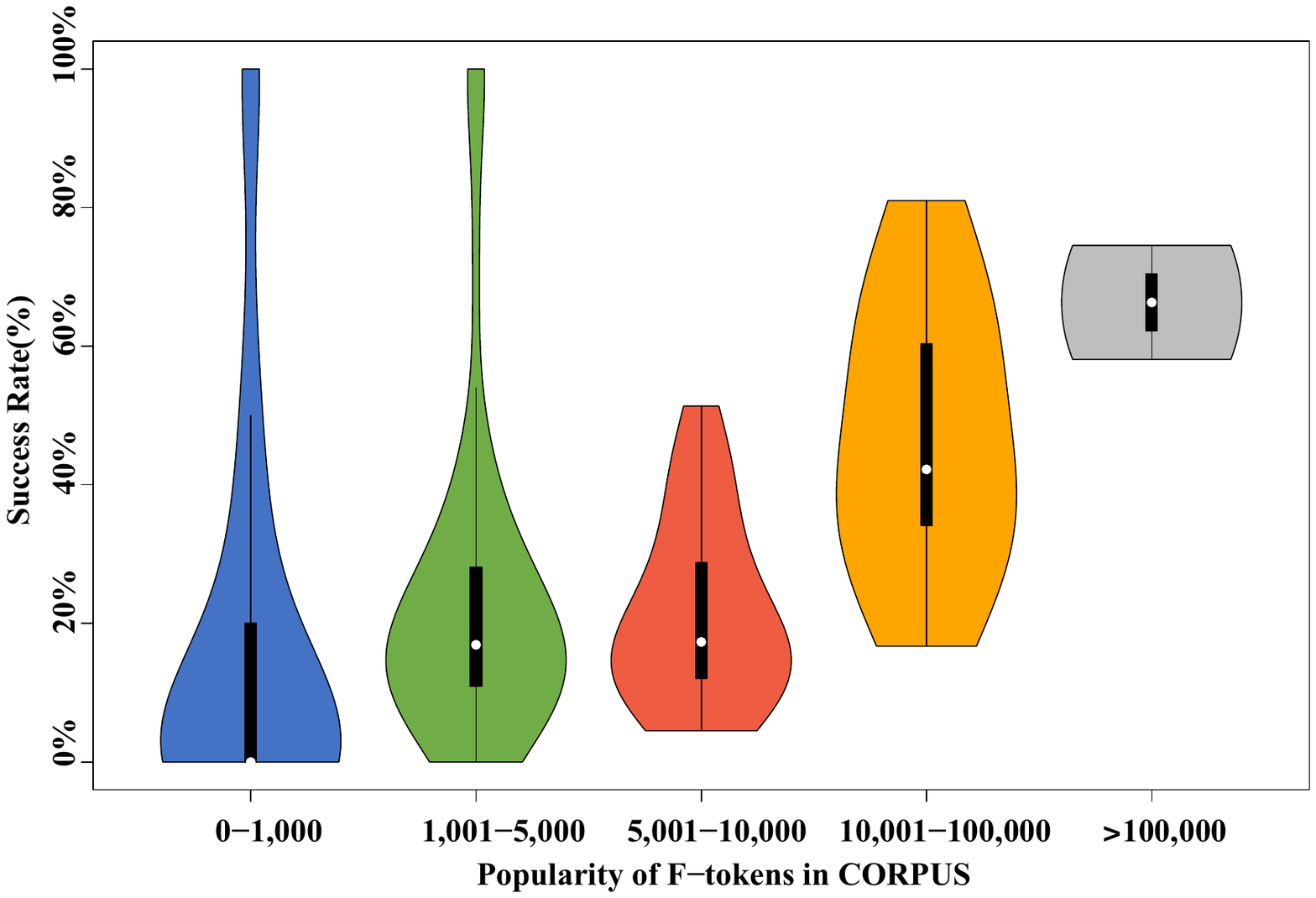}
    \caption{Popularity of \emph{F-token}s  \textbf{ vs. }  Success Rate ($IRMCC$).}
    \label{fig:Distribution19}
\end{figure}

To further validate the correlation between the popularity of \emph{F-token}s and \emph{success rate}, we present their relation in Fig.~\ref{fig:Distribution19}. The horizontal axis specifies the frequency of the \emph{F-token} in the training dataset (\emph{CORPUS\_CP}). The vertical axis specifies the percentages of the method names beginning with the given \emph{F-token} are identified correctly. From this figure, we observe that the \emph{success rate} increases substantially with the increase of \emph{F-token}s' popularity. To validate the statistical significance of the correlation, we conduct the Spearman correlation~\citep{artusi2002bravais} for the two factors. Our results (p-value = 3.4E-69, rho = 0.4)  suggest that they do have a positive correlation, which statistically validates our observation.

Based on the preceding analysis, we conclude that $IRMCC$ works well on the methods whose names start with popular \emph{F-token}s in \emph{CORPUS\_CP}.

\subsection{Method Body Complexity Matters} \label{subsec:LOCAnalysis}

From the method body perspective, we measure the complexity of method bodies by two widely used metrics, i.e., LOC and McCabe's Cyclomatic Complexity~\citep{mccabe1976complexity}. We first investigate the LOC of the sampled 383 methods to see how LOC impacts the performance of $IRMCC$. We find that the average LOC of the success and failure methods is 1.8 and 3.8, respectively. To figure out whether the LOC of success and failure methods has a significant difference, we conduct the Wilcoxon-Mann-Whitney u-test (i.e., Wilcoxon rank sum test)~\citep{ott2015introduction} on these two groups of LOC. The results (statistic=-5.2, p-value=2.1E-07) suggest that there is a significant difference between the LOC of success and failure methods, which further indicates that the methods that are predicted incorrectly may have more LOC in their method bodies. We then analyze the LOC of each method in $NaturalData$, the average number of LOC of success and failure methods is 2.0 and 3.2, respectively. The results of conducting the Wilcoxon-Mann-Whitney u-test on $NaturalData$ (statistic=-80.4, p-value=0.0) further validate that the more LOC a method has, the more likely it will be predicted incorrectly by $IRMCC$.

We then investigate the complexity of method bodies by McCabe's Cyclomatic Complexity which measures the code complexity better. Intuitively, we analyze the code structure of the sampled 383 methods because methods with more loop structures and branched structures tend to be more complicated. We categorize the sampled methods by three basic code structures, i.e., sequential structure, branched structure, and loop structure ordered by complexity, and calculate the average \emph{success rate} of methods with these three structures. If a method consists of more than one structure, we take the more complex one. The results suggest that the average \emph{success rate} of methods with sequential structure, branched structure, and loop structure are 53.0\%, 36.1\%, and 11.1\%, respectively. It indicates that the more complicated the code structure of a method is, the more likely it will be predicted incorrectly by $IRMCC$. To further validate this assumption, we first calculate the McCabe's Cyclomatic Complexity of each method in $NaturalData$, and then conduct the Spearman correlation analysis~\citep{artusi2002bravais} between the McCabe's Cyclomatic Complexity and the \emph{success rate}. Our results (correlation=-0.1, p-value = 0.0) suggest a negative correlation between the two factors, which statistically validates our assumption, i.e., the more complicated a method is, the more likely it will be predicted incorrectly by $IRMCC$. The reason why generation-based approaches work well on simple methods is that simple methods are common in \emph{CORPUS\_CP} and thus easy to retrieve. For methods with complex bodies, it is hard for $IRMCC$ to retrieve similar bodies from \emph{CORPUS\_CP} because these complex methods are rare in \emph{CORPUS\_CP}.

Based on the analysis of methods' complexity, measured by LOC and McCabe's Cyclomatic Complexity, and their \emph{success rate}, we conclude that $IRMCC$ works well on methods with simple bodies, i.e., fewer LOC and lower McCabe's Cyclomatic Complexity while it works less effectively on methods with complex bodies.

\subsection{Requesting for More Advanced Representation Techniques}\label{subsec:SimilarMethodsAnalysis}
Given that $IRMCC$ is based on information retrieval, the retrieved similar methods must embody important information, which may help us figure out why $IRMCC$ frequently fails. As illustrated in Section~\ref{sec:RelatedWork}, when $k=1$, $IRMCC$ retrieves two method names for a test method: one (noted as $mn_2 \in MNs_2$) is the name which is the most similar to the test method name, and the other (noted as $mn_1 \in MNs_1$) is the name of the method whose body is the most similar to the test method's body. $IRMCC$ will regard the test method name as consistent when the \emph{F-token} of $mn_1$ and $mn_2$ are identical. 
We first collect the method names and bodies which are the most similar ones retrieved by $IRMCC$ from \emph{CORPUS\_CP}. Then we compare them with the names and bodies of the target test methods, since the retrieved method names are supposed to have the same \emph{F-token}s as the target test method names in the opinion of Liu et al. 
\begin{table}[]
    \caption{Comparison of the \emph{F-token}s Between Retrieved Methods and Test Methods. }
    \centering
    
    \renewcommand{\arraystretch}{1.2}
    \label{tab:SimilarAnalysis}
    \begin{tabular}{ccc}
    \toprule
                         & \textbf{Name Different}   & \textbf{Body Different} \\ \midrule
    Success Methods      & 2.2\%            & 3.2\%            \\
    Failure Methods      & 38.9\%           & 87.4\%           \\
    All Sampled Methods  & 21.1\%           & 46.7\%           \\ \bottomrule
    \end{tabular}
    \end{table}

Table~\ref{tab:SimilarAnalysis} presents the results of comparing the \emph{F-token}s between retrieved methods and test methods. We analyze the situation of the \emph{F-token}s in success methods, failure methods, and all sampled methods, respectively. The first line in Table~\ref{tab:SimilarAnalysis} presents two possible cases during the comparison. ``Name Different'' means that the \emph{F-token} of $mn_2$ is different from that of the test method name, i.e., the retried method name is not accurate. ``Body Different'' means that the \emph{F-token} of $mn_1$ is different from that of the test method name, i.e., the retried method body is not accurate. The last three lines present the ratios of ``Name Different'' and ``Body Different'' in success methods, failure methods, and all sampled methods, respectively. From Table~\ref{tab:SimilarAnalysis}, we make the following observations:

\begin{itemize}
    \item First, the ratio of ``Name Different'' methods and ``Body Different'' methods is much higher in failure methods than in success methods and even all sampled methods. In addition, ``Body Different'' methods are more than ``Name Different'' methods in success methods, failure methods, and all sampled methods. Especially in failure methods, the ratio of  ``Body Different'' methods is 2.2 = 87.4\%/38.9\% times higher than that of ``Name Different'' methods, which may indicate that retrieving similar method bodies for the test methods is more difficult than retrieving similar method names.
    \item Second, we notice that even in success methods, there are still 2.2\% and 3.2\% of the success methods belonging to the ``Name Different'' and ``Body Different'' categories. This is because the prediction hypothesis of $IRMCC$ only depends on the \emph{F-token}s of $mn_1$ and $mn_2$, and there will be some lucky cases in all success methods. For example, there is a consistent test method $m$ named ``require'', and the \emph{F-token}s of $mn_1$ and $mn_2$ that $IRMCC$ retrieved for $m$ are both ``get''. In this case, $m$ will be considered consistent despite the difference between the \emph{F-token} of $mn_1$ (or $mn_2$) and $m$.
\end{itemize}

\begin{lstlisting}[float,style=JavaStyle,caption={Example For Case 1.},label={lst:example1}]
    // test method
    public GridCacheMvccCandidate removeLock() {
        GridCacheMvccCandidate ret = super.removeLock();
        locPart.onUnlock();
        return ret;
    }
    
    
    // most similar body to the test method body
    protected Control createButtonBar(Composite parent){
        Control control = super.createButtonBar(parent);
        updateWidgets();
        return control;
    }
    \end{lstlisting}
    
    \begin{lstlisting}[float,style=JavaStyle,caption={Example For Case 2.},label={lst:example2}]
    
    // test method
    public void shutdown() {
        closeConnection();
    }
    
    // most similar body to the test method body
    public void close() throws IOException{
        closeConnection();
    }
    \end{lstlisting}
From the observations we made from Table~\ref{tab:SimilarAnalysis}, we can infer that the retrieved method bodies play a more important role than the retrieved method names in checking whether a method name is consistent with its body.
To this end, the first author and the third author independently analyze the 173=198*87.4\% failure cases that are caused by ``Body Different'', categorizing each case into one of the two failure types:
\begin{itemize}
    \item \textbf{Case 1. (82.7\%)} The retrieved method bodies are not functionally similar to those of the test methods. For an example shown in Listing~\ref{lst:example1}, the function of the test method is to remove the lock of the object while the function of the retrieved method body is to create a new button bar object.
    \item \textbf{Case 2. (17.3\%)} The retrieved method bodies are functionally similar to those of the test methods. Listing~\ref{lst:example2} shows an example, where the test method and the retrieved method body both implement the function of closing connection.
\end{itemize}
The Cohen's kappa coefficient of agreement~\citep{cohen1960coefficient} between the two authors is 0.94. For the disagreement cases, the second author acts as an arbitrator to make the final decision. Based on the above two categories, we can infer that there are two corresponding reasons leading to the failure of $IRMCC$. In the first case, 82.7\% of failure methods are caused by the poor method body representation because the retrieved method bodies are not similar to those of the test methods. Subsequently, more advanced method representation techniques, e.g., CodeT5~\citep{wang2021codet5}, CodeBERT~\citep{feng2020codebert}, should be exploited to improve the performance of IR-based approaches. In the second case, the hypothesis that methods implementing similar behavior in their body code are likely to be consistently named with similar names does not hold sometimes. As a result, 17.3\% of methods fail due to violation of the hypothesis. This case exists because different developers have their naming habits and thesaurus. Even for the same method, developers may name it differently.

\begin{tcolorbox}
    \textit{Summary:}
    $IRMCC$ works well on the methods whose names are popular in \emph{CORPUS\_CP} and bodies can be found in \emph{CORPUS\_CP}. It also works well on the methods with limited LOC and low McCabe's Cyclomatic Complexity because these methods share some easy-to-find structural characteristics. It fails briefly because the method body representation is not efficient and the hypothesis, i.e., two methods with similar bodies should have similar names, does not hold all the time. More advanced method representation techniques should be exploited to improve the performance.
\end{tcolorbox}

\section{RQ4: Analysis of Generation-based Approaches}\label{sec:RQ4}

In this research question, we investigate where and why the generation-based approaches, i.e., $CAN$, $MN_{IRE}$, $Cognac$, and $GTNM$, work or fail. Following the patterns in Section~\ref{sec:RQ3}, we also analyze where and why these four DL-based approaches work or fail from both method name perspective and method body perspective. The sampled data is the same 383 methods adopted in Section~\ref{sec:RQ3}. From the method name perspective, we analyze the threshold of similarity between generated names and test names. From the method body perspective, we also analyze the method body complexity by measuring LOC and  McCabe's Cyclomatic Complexity of methods. We also conducted a qualitative analysis to further explore the failure reason. At last, based on the above analysis, we propose two possible ways to further improve the performance of identifying inconsistent method names.\\

\subsection{Method Name's Similarity and Length Matter} \label{subsec:ThresholdAnalysis_20}

\begin{figure*}[htb]
    \centering
        \subfigure[Methods' Similarities ($CAN$).]{
        \label{fig:subfig:SampleSimilarity_CAN}
    \includegraphics[width=0.4\linewidth ,trim={0 40 0 40 }, clip]{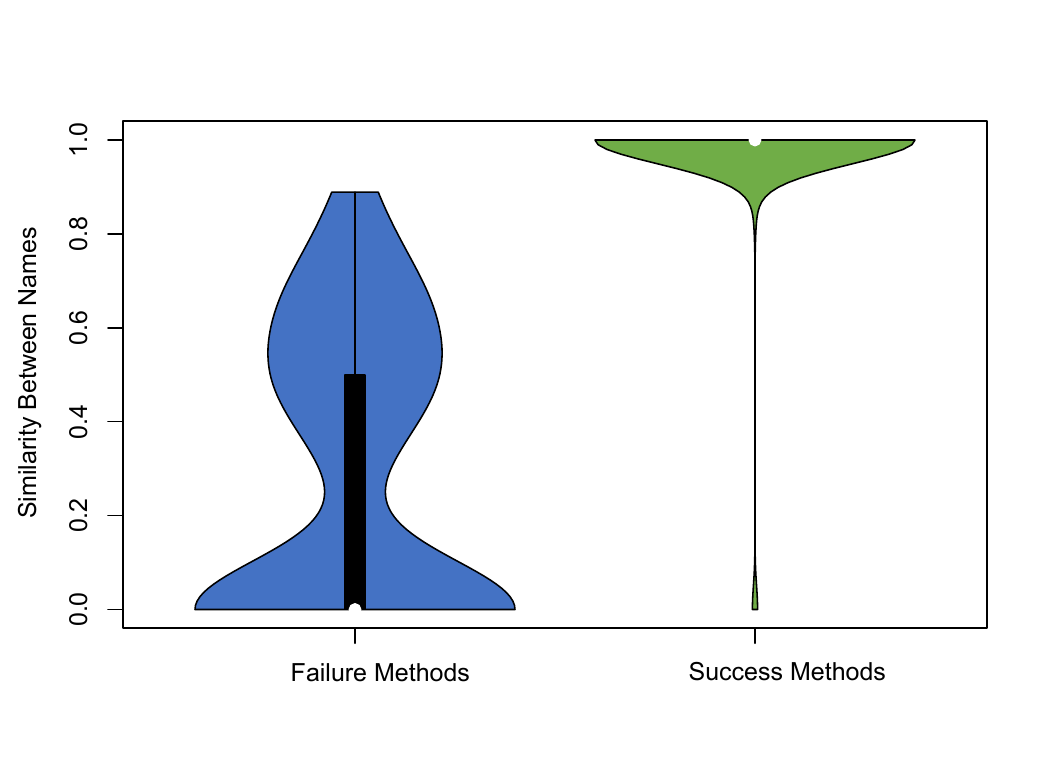}}
    \hspace{0.01\linewidth}
    \subfigure[Methods' Similarities ($MN_{IRE}$).]{
        \label{fig:subfig:SampleSimilarity_MNIRE}
    \includegraphics[width=0.4\linewidth ,trim={0 40 0 40 }, clip]{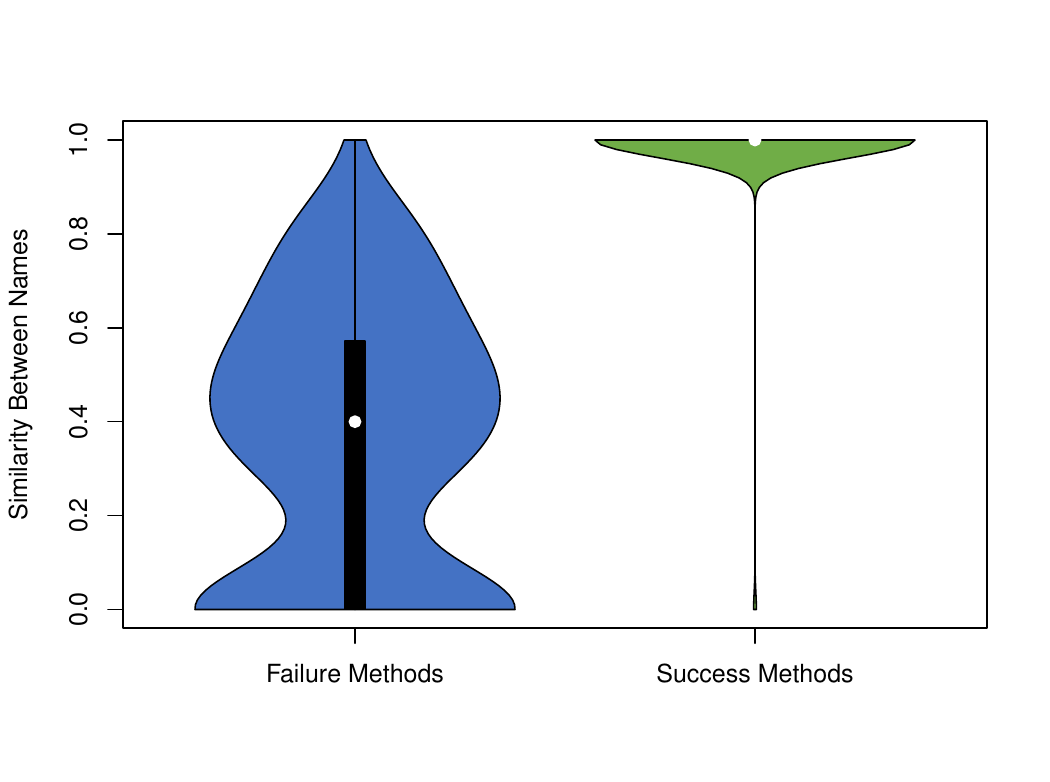}}
    
    \vfill 

    \subfigure[Methods' Similarities ($Cognac$).]{
        \label{fig:subfig:SampleSimilarity_Cognac}
    \includegraphics[width=0.4\linewidth ,trim={0 40 0 40 }, clip]{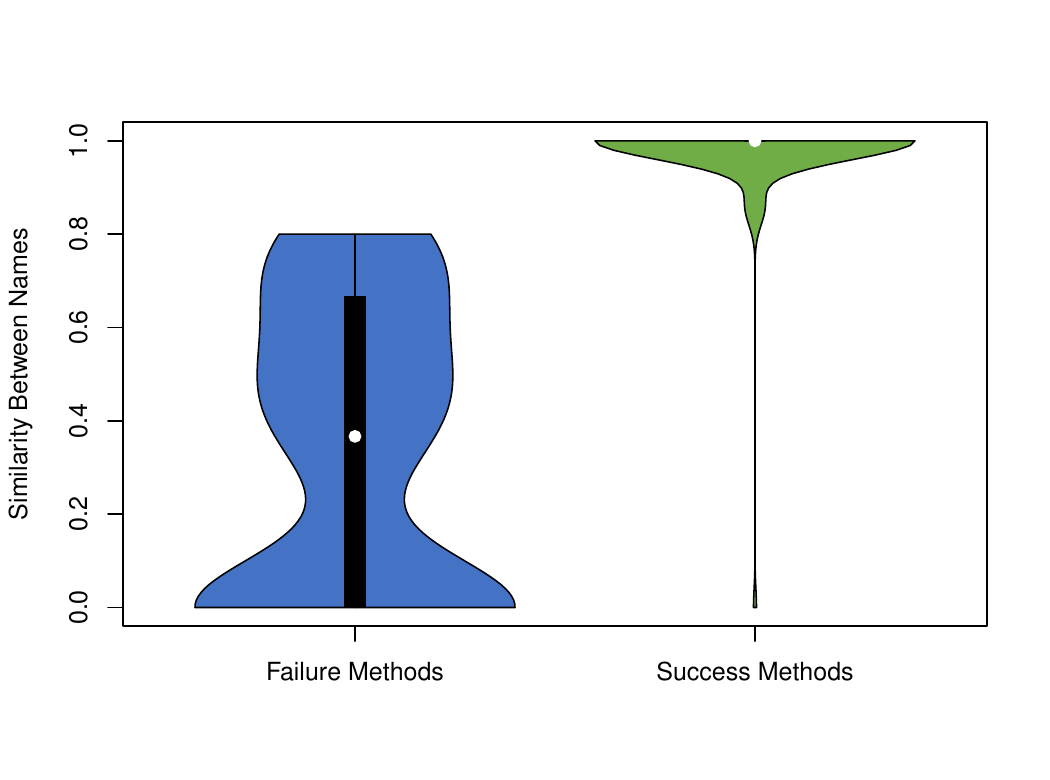}}
       \hspace{0.01\linewidth}
    \subfigure[Methods' Similarities ($GTNM$).]{
        \label{fig:subfig:SampleSimilarity_GTNM}
    \includegraphics[width=0.4\linewidth ,trim={0 40 0 40 }, clip]{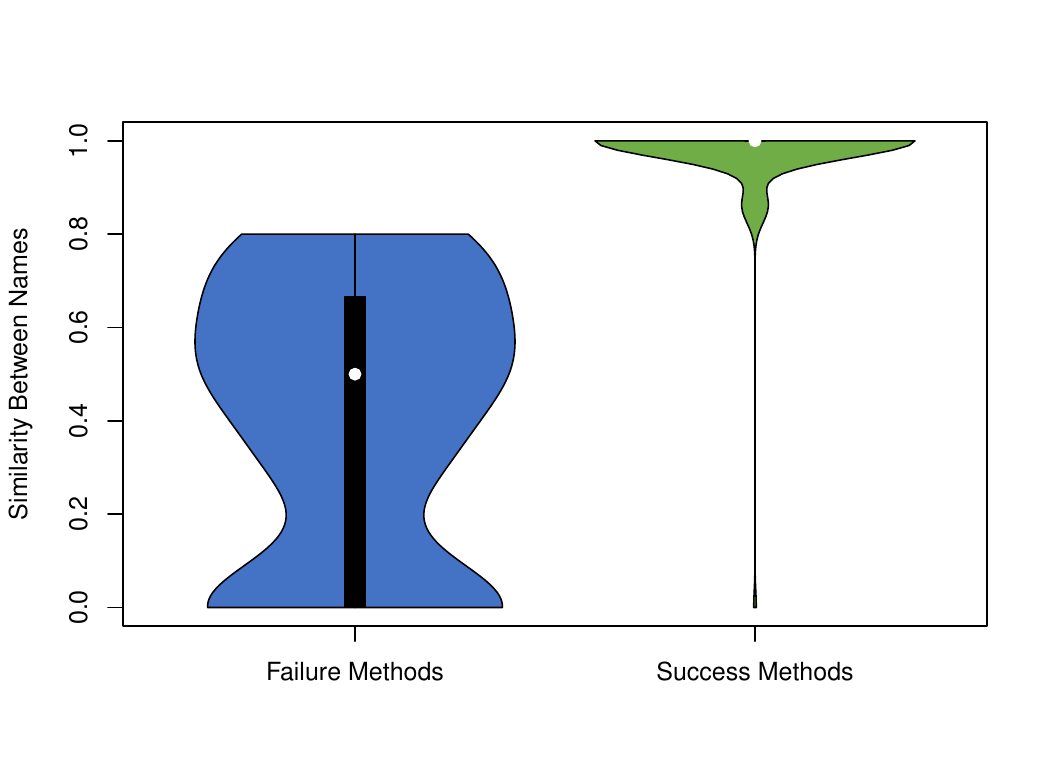}}
    \caption{Distribution of Methods Similarities.}
    \label{fig:Similarity}
\end{figure*}

From the method name perspective, we analyze how the similarity between generated method names and test method names impacts the performance of generation-based approaches, i.e., $MN_{IRE}$, $CAN$, $Cognac$, and $GTNM$. The distribution of the similarity between success method names and their corresponding generated names, and the similarity between failure method names and their corresponding generated names are presented in Fig.~\ref{fig:subfig:SampleSimilarity_CAN}, Fig.~\ref{fig:subfig:SampleSimilarity_MNIRE}, Fig.~\ref{fig:subfig:SampleSimilarity_Cognac}, and Fig.~\ref{fig:subfig:SampleSimilarity_GTNM}. We can see that in the success methods, the distribution of similarities between success method names and their corresponding generated names is funnel-like (right part of each figure). The majority of the similarity of success methods is 1.0, which means that only if the generated names share all the sub-tokens with the ground truth names can the consistent test methods be identified correctly. This also indicates that high \emph{threshold} represents the generated names should be perfect since the token number of method names is usually less than five~\citep{DBLP:conf/icse/AlsuhaibaniNDCM21}. In addition, as we can see from the distributions of similarities between failure method names and their corresponding generated names (left part of each figure), there are even some failure methods with high similarities over 0.8, which means that in these cases only one sub-token is different from the ground truth names. The above analysis suggests that the similarity calculation method that only considers the lexical information of method names is not sound enough. Furthermore, we also find some generated names that are synonyms to the ground truth names, e.g., ``accept'' (ground truth: ``receive'') and ``get'' (ground truth: ``retrieve'').  This indicates that future generation-based approaches should consider measuring the method name similarity not only lexically but also syntactically and semantically.

We also analyze how the length of method names (i.e., the number of sub-tokens in a method name) impacts the performance of generation-based approaches.  For  $CAN$, $MN_{IRE}$, $Cognac$, and $GTNM$, we calculate the length of all the involved methods in $NaturalData$ and conduct the Spearman correlation analysis~\citep{artusi2002bravais} between the length of method names and their success rate.
The results (statistic=-0.01, p-value=1.3E-07 for $CAN$, statistic=-0.30, p-value=0.0 for $MN_{IRE}$, statistic=-0.18, p-value=9.4E-293 for $Cognac$, and statistic=-0.10, p-value=1.4E-225 for $GTNM$) suggest that there is a weak negative relationship between these two factors. This may be because all generation-based approaches generate the sub-tokens of method names one by one, making it more challenging to generate longer method names (with more sub-tokens) correctly compared to shorter ones. The relationship is weak because most of the method names are of medium length (2-5 sub-tokens), and the extremely long method names are rare.

\subsection{Method Body Complexity Matters}\label{subsec:LOCAnalysis_20}
\begin{figure}
    \centering
    \includegraphics[width=\textwidth ,trim={40 80 40 80 }, clip]{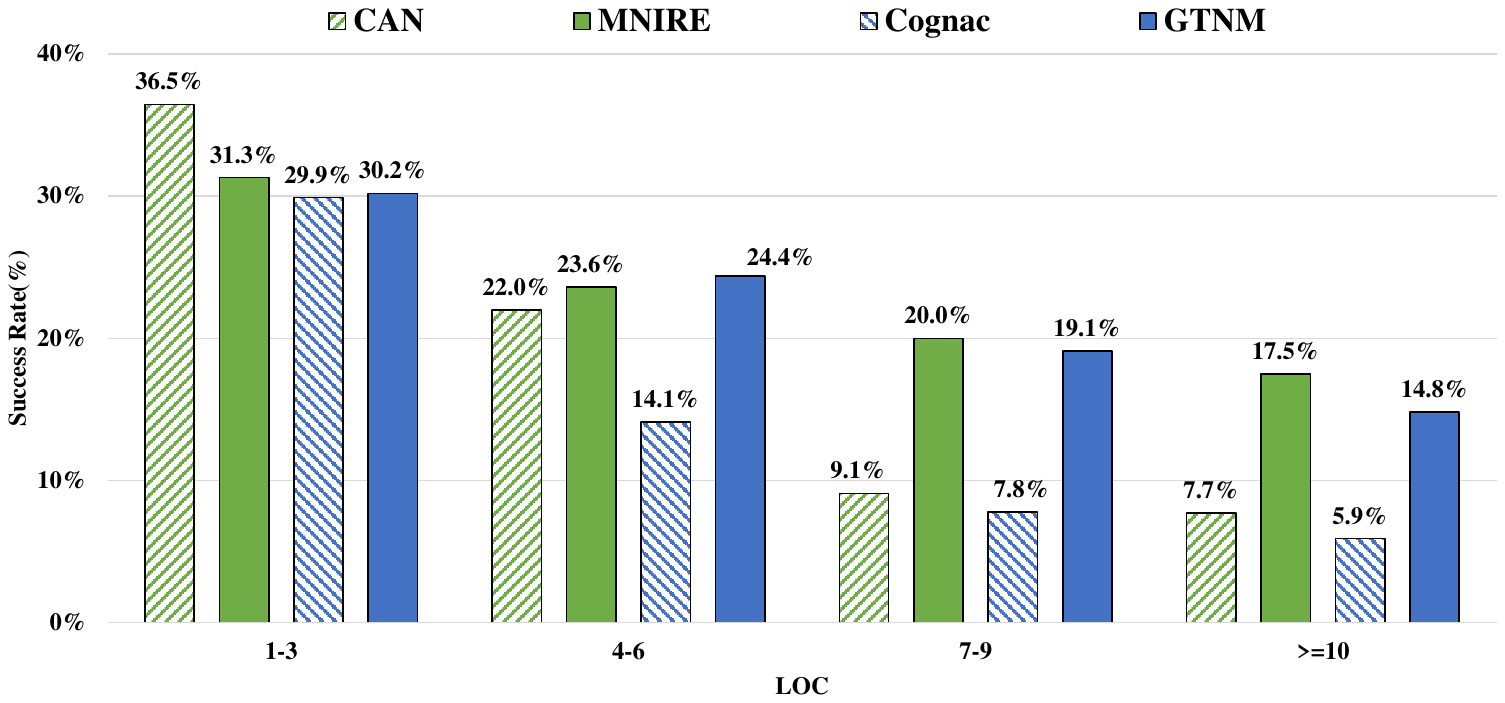}
    \caption{LOC \emph{VS.} Success Rate.}
    \label{fig:LOCAnalysis}
\end{figure}

To investigate how the method body complexity impacts the generation-based approaches, we first analyze the lines of code of all the methods in $NaturalData$ (LOC of methods in success ones and failure ones are analyzed separately). Then we manually segment LOC into four groups and explore the average \emph{success rate} of each group. The analysis results are presented in Fig.~\ref{fig:LOCAnalysis}. The horizontal axis denotes the LOC, and the vertical axis denotes the \emph{success rate} of methods. From Fig.~\ref{fig:LOCAnalysis}, we observe that with the increase of LOC, the \emph{success rate} of methods in $NaturalData$ decrease significantly. 
Similar to IR-based approaches, the more LOC a method has (i.e., more complex), the less likely the generation-based approaches will predict it correctly. The results ( correlation=-0.32, p-value=7.9E-312 for $CAN$, correlation=-0.21, p-value=2.3E-22 for $MN_{IRE}$, correlation=-0.22, p-value=3.7E-71 for $Cognac$, and correlation=-0.39, p-value=5.2E-210 for $GTNM$) of Spearman correlation analysis between LOC and \emph{success rate} further validate this observation.

We then calculated McCabe's Cyclomatic Complexity of all the methods in $NaturalData$ to see how the complexity of methods impacts the performance of $CAN$, $MN_{IRE}$, $Cognac$, and $GTNM$. After that, we conducted the Spearman correlation analysis between McCabe's Cyclomatic Complexity of methods and their \emph{success rate}. Our results (correlation=-0.21, p-value=1.0E-33 for $CAN$, correlation=-0.34, p-value=1.6E-381 for $MN_{IRE}$, correlation=-0.45, p-value=5.6E-210 for $Cognac$, and correlation=-0.27, p-value=5.7E-129 for $GTNM$)  suggest that they are indeed negatively correlated, which statistically validates our assumption. This result also indicates that the more complicated a method is (i.e., more complex), the more likely it will be predicted incorrectly by generation-based approaches.

From the analysis of LOC and McCabe's Cyclomatic Complexity, we conclude that generation-based approaches, i.e.,$CAN$, $MN_{IRE}$, $Cognac$, and $GTNM$, work well on methods with simple bodies, i.e., fewer code lines and easier structure. The reason might be that simple methods tend to have less number of tokens in method bodies and thus easy for deep learning models to extract information and generate the correct sub-tokens.

\subsection{Qualitative Analysis}
To further explore the failure reason of generation-based approaches and move beyond the purely historical view of naming based on past data, we conducted a qualitative analysis to investigate the names generated by generation-based approaches. 

Since the SOTA generation-based approach is $GTNM$, and it achieves the best overall performance in identifying inconsistent method names in $NaturalData$, we only analyzed the generation results of $GTNM$ to investigate the reason why generation-based approaches fail. In addition, Liu et al.~\citep{liu2022learning} have already explored the cases where $GTNM$ cannot make a correct recommendation in their paper, so we focus on its ability to identify inconsistent method names instead of name recommendation in this qualitative analysis. Consequently, we sampled 332 inconsistent method names from all the inconsistent method names (2,443) in $BenMark$ with a confidence level of 95\% and a margin of error of 5\%. This ensures that our sample is representative and reliable to reflect the characteristics of all the inconsistent methods. In the 332 samples, $GTNM$ failed to identify 18 of them and succeeded in identifying 314 of them.
Recall that the rationale of generation-based approaches is to generate a name first and then compare it with the name under test. Consequently, triplets like $<BuggyName, FixedName, MethodBody>$, and the names generated by $GTNM$ (we call them $GeneratedName$), are taken for reference. 

The first and second authors manually inspected these 332 cases independently. The major focus was on which name ($GeneratedName$ or $FixedName$) was better, and why would $GTNM$ generate these names. The two authors discussed the disagreements until they reached a consensus. The Cohen's kappa coefficient is 0.79, indicating a substantial agreement between the two authors.

Finally, we made the following observations:
\begin{itemize}
    \item First, in 18 failure cases, 88.8\%(=16/18) false negatives are due to the inability of $GTNM$ to handle \textbf{``Narrow''} type, i.e., $GeneratedName$ (or $BuggyName$) lacks some additional information compared to $FixedName$. An example is presented in the below code snippets. The $GeneratedName$ (or $BuggyName$) is \textbf{``getConnectionFactory''}. However, developers think that \textbf{``getOrCreateConnectionFactory''} is a more descriptive and appropriate name for the method body.
    \item Second, in 314 success cases, there are only 19.7\% (=62/314) $GeneratedName$s that are identical to $FixedName$s. That is to say, in these cases, the generated names coincided with the intentions of developers. Although the remaining 80.3\% (=252/314) cases are still identified correctly, these $GeneratedName$s are not identical to $FixedName$s. Consequently, we further analyze these cases.
    \item Third, in the 252 correctly identified cases, we found that in 80.2\% cases, $FixedName$s are better than $GeneratedName$; In 16.6\% cases, $FixedName$ and $GeneratedName$ are both appropriate for the method body; In 3.2\% cases, $GeneratedName$ is even better than $FixedName$, Additionally, in the 80.2\% poorly generated cases, 22.3\% cases are due to the lack of additional information, i.e.,  lack of several sub-tokens compared to $FixedName$s. This further indicates that existing generation-based approaches struggled to identify inconsistent method names that only lack some additional information (i.e., struggle to handle \textbf{``Narrow''} type). 12.9\% cases are because of the inability to handle acronyms. The above findings to some extent coincide with the conclusions reached by Liu et al.~\citep{liu2022learning}.
\end{itemize}

\begin{lstlisting}[float,style=JavaStyle,caption={Examples of False Negatives.},label={FalseNegatives}]
// GTNM generated "getConnectionFactory" for the below method body.
public ConnectionFactory getOrCreateConnectionFactory() {
    if (connectionFactory == null) {
        connectionFactory = createConnectionFactory();
    }
    return connectionFactory;
}
    \end{lstlisting}

From the above findings, we conclude that the ability of generation-based approaches still needs to be improved to more accurately identify inconsistent method names, especially in identifying the ones belonging to \textbf{``Narrow''}. The likely reason is that in the \textbf{``Narrow''} case, the inconsistent method name encompasses the majority of the functionality of the method bodies. This requires generation-based approaches to thoroughly understand the functionality of the method bodies and generate more descriptive (i.e., longer) names, which is quite challenging.

\begin{tcolorbox}
    \textit{Summary:}
        Generation-based approaches, i.e.,$CAN$, $MN_{IRE}$, $Cognac$, and $GTNM$ work well on methods with short names, few LOCs, and low complexity. Generation-based approaches fail because of the less effective name similarity calculation method. They particularly struggle to identify inconsistent method names of \textbf{``Narrow''} type. 
\end{tcolorbox}

\section{Discussion}\label{sec:Discussion}
\subsection{Threats to Validity}
We now discuss the threats to the validity of our study, following common guidelines for empirical studies~\citep{kitchenham2002preliminary}.
\subsubsection{Threats to External Validity}
A threat to external validity is the limited size and diversity of the dataset employed in the evaluation. Given the limited size and diversity of the dataset, it is possible that our findings cannot be generalized anywhere. Notably, we reuse the subject applications selected and extensively used by existing research~\citep{Liu2019,Nguyen2020,Li2021,wang2021lightweight}. Reusing such well-known open-source applications facilitates third-party replication of our empirical study. 

\subsubsection{Threats to Constructive Validity}\label{subsubsec:Threats}
A threat to constructive validity is that the labels of the items (i.e., whether a given method name is consistent with its method body) could be inaccurate. Following Liu et al.~\citep{Liu2019}, we label the method names by mining version control systems.  However, this method can not guarantee that the method name changes are associated with the inconsistency between names and bodies. To reduce the false positives, we added a manual inspection after the automatic identification. Three developers are invited to rate the method names, and a judgment criterion has been summarized. Then two authors manually inspected the dataset with this criterion, aiming to minimize the number of false positives. In addition, it is infeasible to manually inspect the consistent method names due to their overwhelming number. To figure out the quality of labels on consistent method names, We sampled parts of the consistent methods and examined the false positive rate. The results suggest that the false positive rate is only 4.4\%.

\subsubsection{Threats to Internal Validity}
A threat to internal validity is that we (instead of the original authors) calibrate the evaluated approaches which may not result in the optimal settings. All the evaluated approaches are learning-based and contain a few parameters that should be optimized according to the given data. Although we have tuned such parameters following the original paper, likely, we may not find the optimal settings because of a lack of deep understanding of their models and implementations. To facilitate the replication, we specify how we tune the parameters in Section~\ref{subsec:tuning}, and make the whole replication package publicly available at GitHub~\citep{MCC}.

\subsection{Limitations}
\label{sub:limitation}
The first limitation of the paper is that it is confined to Java methods only. Both of the evaluated approaches should apply to source code in various programming languages because they do not depend on the special characters of a given programming language. However, their public implementations are currently confined to Java. To this end, our evaluation involves Java code only.  In the future, it could be interesting to investigate how well such DL-based approaches work on other programming languages, e.g., C++, JavaScript, and Python.

The second limitation is that we do not evaluate the state-of-the-art DL-based approaches in the field. Instead, we identify testing items by mining the commit histories of open-source applications. However, inconsistent names accounted for in the field could be significantly different from those recorded in the version control systems. Many inconsistent method names have likely been renamed before they are committed and merged into the main branches in the version control systems. Consequently, they could be missed by the mining-based approach. In the future, we should evaluate the state-of-the-art DL-based approaches in the field by collecting feedback from developers.

\subsection{Complementary Analysis}\label{sub:ComplementaryAnalysis}

    \begin{figure}[]
        \centering
        \includegraphics[width=0.75\linewidth ,trim={0 60 0 60 }, clip]{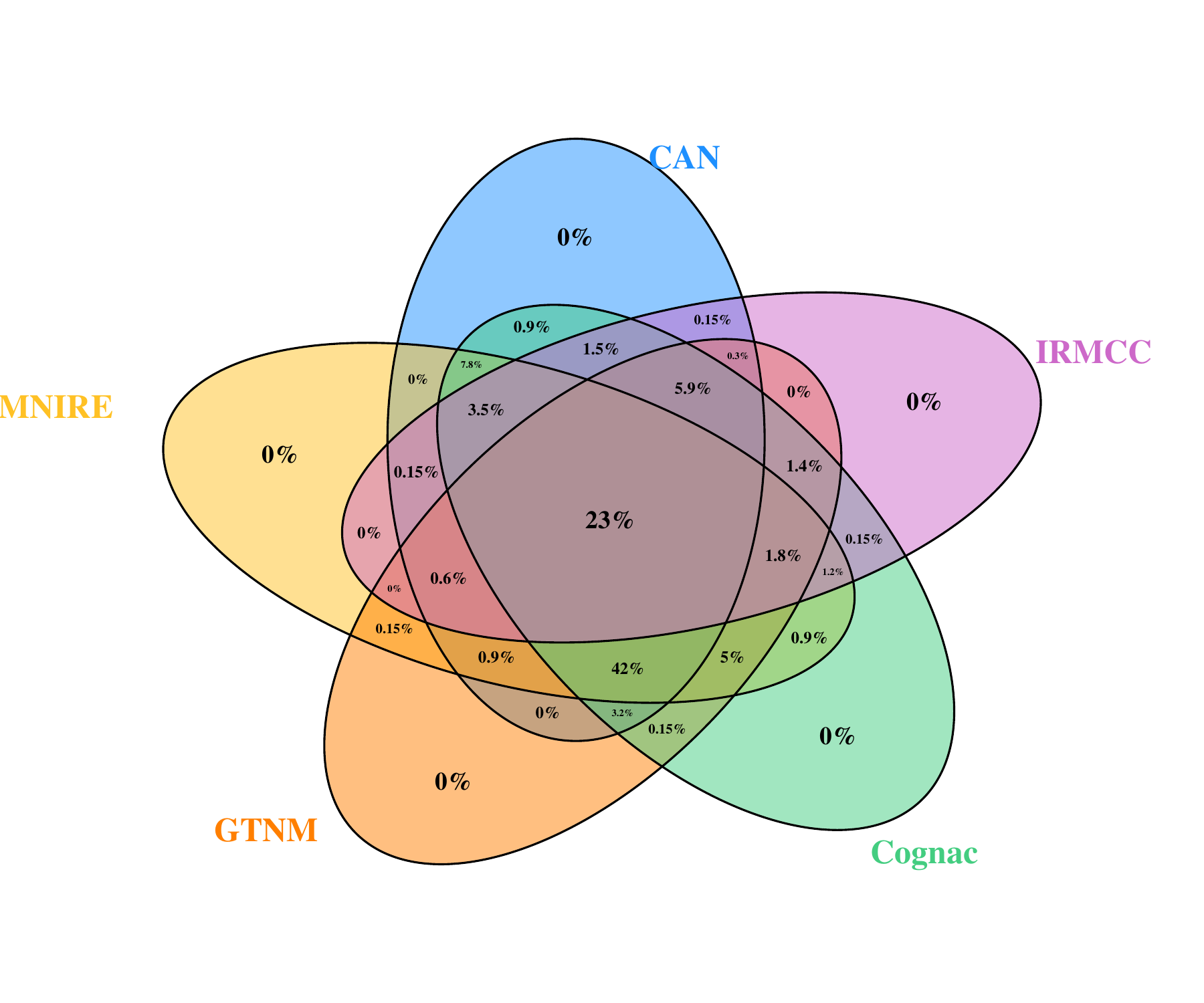}
        \caption{Complementarity of Five Selected Approaches in Identifying Inconsistent Method Names.}
        \label{fig:Complementarity}
    \end{figure}
    In this section, we further explored the complementarity of five selected approaches. We analyzed their performance in identifying inconsistent method names and drew a Venn diagram using VennDiagram~\citep{VennDiagram} package in R language.
    Analysis results are shown in Fig.~\ref{fig:Complementarity}. From the above figure, we make the following observations:
    \begin{itemize}
        \item There are 23.0\% of inconsistent method names that can be commonly identified by $IRMCC$ and $CAN$, $MN_{IRE}$, $GTNM$, and $Cognac$. Notably,  not a single inconsistent method name can be identified exclusively by any of the five approaches. This pattern is also observed in the four generation-based approaches: only 0.15\% inconsistent method names can be identified exclusively by $CAN$ and $Cognac$, respectively. This suggests that there is little complementarity among the five approaches in identifying inconsistent method names.  
        \item  The above findings are reasonable when analyzed from the aspect of performance. since the precision of identifying inconsistent method names is significantly affected by the number of consistent method names in testing data. Consequently, we only consider the recall of identifying inconsistent method names. As shown in Table~\ref{tab:Comparison19_real} - Table~\ref{tab:comparison_Cognac_real}, the recall with $IRMCC$ is only 65.0\%. In contrast, the recall of the generation-based approaches is 94.9\%, 87.8\%, 83.7\%, and 93.3\%. The generation-based approaches can almost entirely cover the instances identified by the IR-based approach. Furthermore, due to the near-perfect recall of generation-based approaches, very few instances are identified exclusively by them. 
        \item The above findings also coincide with the conclusions reached in Section~\ref{sec:RQ3} and Section~\ref{sec:RQ4}. Both IR-based approaches and generation-based approaches are good at identifying methods with short bodies and simple names while having difficulty in identifying complex ones. For example, if a method is as simple as a getter method (see Listing~\ref{ThreeTypicalExamples}), it is easy to retrieve from the code corpus and easy to generate from scratch. Otherwise, if it is an extremely complex method, it is hard to retrieve a similar method from the code corpus and hard to generate from scratch because the semantics could also be complicated.
    \end{itemize}
    
\subsection{Possible Ways to Improve The Performance}
\subsubsection{Leveraging Contrastive Learning}\label{subsubsec:ContrastiveLearning}

\begin{figure}
    \centering
    \includegraphics[width=\textwidth ,trim={0 0 0 0 }, clip]{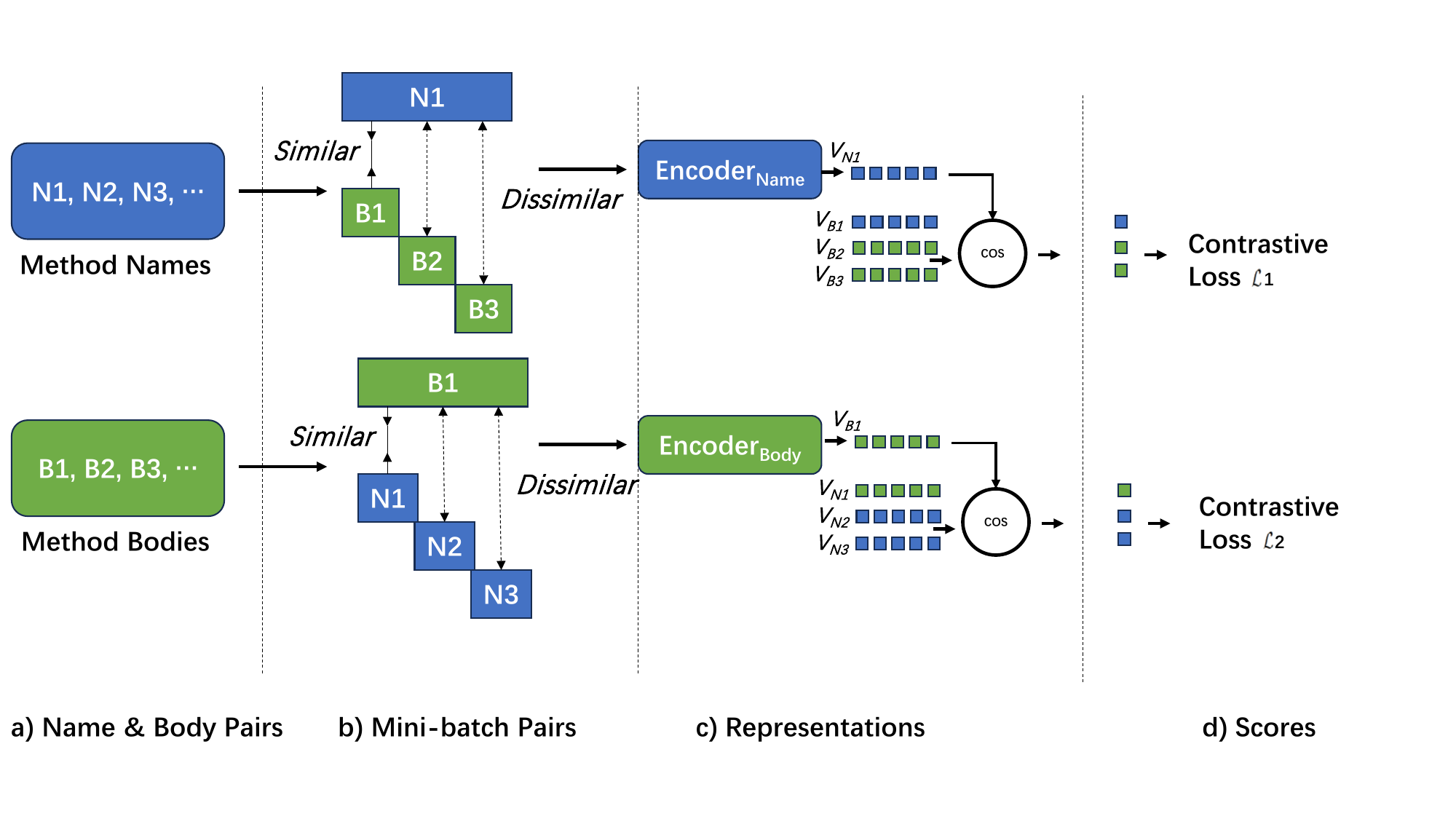}
    \caption{Overview of the Contrastive Pre-training.}
    \label{fig:ContrastiveLearning}
\end{figure}

Although the mainstream (SOTA) approach is based on name generation, it is an indirect solution for the identification of inconsistent method names, as it was originally designed to generate names for method bodies.  Furthermore. In addition, as analyzed in Section~\ref{subsec:ThresholdAnalysis_20}, generation-based approaches struggled to identify inconsistent method names of \textbf{Narrow} type, these approaches are heavily impacted by subsequent similarity calculation metrics. For the above reasons, its performance in identifying inconsistent method names is substantially constrained by the effectiveness of method name recommendation. Consequently, based on the analysis presented in the previous sections, we propose a more efficient method specially designed to identify inconsistent method names. 
Conceptually, contrastive learning leverages encoder networks (could be any type of structures, e.g., RNN, LSTM, Transformer) to convert the instances (in this paper, method names and bodies) into vector representations, aiming to pull close the distance between similar instances while pushing away the distance between dissimilar instances~\citep{oord2018representation}.  Contrastive learning has been extensively leveraged to represent the semantic meaning of identifiers~\citep{Chen2022} and source code~\citep{Ding2022}.  The overview of the contrastive pre-training is presented in Fig.~\ref{fig:ContrastiveLearning}. The training process is self-supervised. Since the majority of method names and bodies are consistent with each other, the training corpus can be easily obtained. The input instances are natural pairs of method names and bodies. At each training step, a mini-batch contains pairs of names and bodies. The training aims to minimize the distance within a pair, e.g., $<N_1,B_1>$, while maximizing the distance of other pairs, e.g.,  $<N_1,B_2>$ and $<N_1,B_3>$. To achieve this, $Encoder_{Name}$ and $Encoder_{Body}$ are adopted to convert names and bodies into vector representations. Note that the $Encoder_{Name}$ and $Encoder_{Body}$ can be initialized by the weights of trained models, e.g., CodeBert~\citep{feng2020codebert}.  Contrastive loss $\mathcal{L}_1$ and $\mathcal{L}_2$, which are InfoNCE~\citep{oord2018representation}, are calculated based on the cosine similarity between the encoded vector representations. Finally, $Encoder_{Name}$ and $Encoder_{Body}$ are optimized with the gradient descent, respectively. After the pre-training, we will obtain two encoders for method names and bodies, i.e., $Encoder_{Name}$ and $Encoder_{Body}$. Then in the downstream tasks, i.e., identification of inconsistent method names, names and bodies are encoded into vectors, respectively. Finally, these two vectors can be concatenated and fed into a softmax layer to achieve a binary classification. We can fine-tune the model with the manually inspected data in $BenMark$, achieving better performance.

The advantage of contrastive learning is two-fold:
On the one hand, traditional IR-based approaches heavily rely on similarity calculation metrics. However, the retrieval based on similarity calculation from a large code corpus could be ineffective sometimes. This inference can be supported by our analysis in Section~\ref{subsec:SimilarMethodsAnalysis}. 82.7\% of the failed samples are due to the dissimilarity between method bodies. In addition, it is significantly constrained by the quality of the constructed code corpus. By contrast, contrastive learning incorporates similarity calculation in the training process, solving the disadvantages of IR-based approaches.
On the other hand, the traditional generation-based approaches heavily rely on per-training tasks for the alignment between natural language and programming language. It also suffers from the problem of similarity calculation. In addition, it requests large amounts of data for training. However, self-supervised contrastive learning requires less data for training, solving the disadvantages of generation-based approaches.

\subsubsection{Leveraging The Ability of LLM}\label{subsubsec:LLM}
To the best of our knowledge, there are no LLM-based approaches for now to identify inconsistent method names.
The LLMs like GPT4 have been proven to achieve great performance in a series of software engineering tasks~\citep{noever2023can,dong2023self,xia2023keep}.
We believe that the LLMs will also have great potential in solving the problem of identification of inconsistent method names. With the emergent ability and proper prompts, LLMs are likely to thoroughly understand the functionality
of the method bodies and identify inconsistent method names more accurately.

Recently, LLMs are revolutionizing the software engineering field by acting as intelligent agents that enhance productivity and streamline various processes~\citep{xia2024agentless, Hong2024}. It is also interesting to explore how to develop one or multiple agents for the identification of inconsistent method names.

\subsection{Take-away Messages}
\label{sub:take-away}
\textbf{1. Inconsistent Method Names May Be Renamed by Only Changing Few Sub-tokens}
As is presented in Section~\ref{subsubsec:ConstructionOfBenMark}, in real scenarios, developers may only change one or several sub-tokens of the inconsistent method names with the most sub-tokens remaining. Among the inconsistent method names in $BenMark$, 42.1\% inconsistent method names are of \textbf{``Narrow''} (34.3\%) or \textbf{``Generalize''} (7.8\%) type. That is to say, 42.1\% inconsistent method names are renamed by adding or removing one (in most cases) or multiple sub-tokens. The remaining 57.9\% inconsistent method names are of \textbf{``Change''} type. Consequently, the implication is that since almost half of the inconsistent method names may just be inconsistent with their method bodies in one sub-token, it should be noticed during the design of the approaches.
 
\textbf{2. Two Possible Ways for Better Identifying Inconsistent Method Names}
As is illustrated in Section~\ref{subsubsec:ContrastiveLearning}, the advantages of contrastive learning are two-fold:
On the one hand, traditional IR-based approaches heavily rely on retrieval based on similarity calculation from a large code corpus, which could be ineffective sometimes. In addition, it is significantly constrained by the quality of the constructed code corpus. By contrast, contrastive learning incorporates similarity calculation in the training process, solving the disadvantages of IR-based approaches.
On the other hand, the traditional generation-based approaches request large amounts of data for training. However, self-supervised contrastive learning requires less data for training, solving the disadvantages of generation-based approaches. As is illustrated in Section~\ref{subsubsec:LLM}, the ability of LLM has great potential in improving the performance of identification of inconsistent method names. 

\textbf{3. Proportions of Positive and Negative Items in Testing Data Have Great Impact on the Performance of DL-based Approaches.} As is presented in Section~\ref{sec:RQ2}, changing the ratio of inconsistent and consistent method names substantially impacts the performance of the evaluated approaches. This indicates that we should evaluate approaches to testing datasets that are constructed according to real scenarios instead of intentionally controlled balanced ones.  

\textbf{4. Performance of Existing DL-based Approaches Needs to Be Improved Before Being Applied to the Industry.} 
The performance of all the five evaluated approaches on testing data with a natural ratio of inconsistent and consistent methods is not satisfying (the precision is less than 1\%). It means that there are overwhelming false positives, and they will confuse users when using the approaches. Given the importance of having a consistent method name, future studies should be dedicated to improving the current performance of identifying inconsistent method names.

\textbf{5. Multiple Benchmarks Should Be Used to Fully Evaluate The Performance of DL-based Approaches.} Although the dataset that is close to the real scenarios is of great value in reflecting the real performance of DL-based approaches, the balanced dataset may also to some extent exhibit the ability to identify inconsistent method names of DL-based approaches. Consequently, we should try to use multiple benchmarks to fully evaluate the performance of DL-based approaches.

\section{Conclusion and Future Work}\label{sec:Conclusion}
Inconsistent method names could be misleading and result in serious software defects. Consequently, a few automated approaches have been proposed to identify such misleading method names. However, we still lack a comprehensive evaluation of these state-of-the-art approaches in a more natural dataset, and where and why these approaches work or fail are still unclear. This paper selects five state-of-the-art DL-based approaches in this field (one IR-based approach and four generation-based approaches) and evaluates them on our constructed large test dataset that is clean, manually inspected, and close to real-world scenarios. The evaluation results suggest that the ratio of inconsistent and consistent methods in testing data impacts the performance substantially. However, switching the within-project setting to a widely-used cross-project setting slightly changes the performance of the evaluated approaches. Our analysis results suggest that there are two major reasons why IR-based approaches fail: 1) the adopted method body representation technique is not efficient enough, and 2) the hypothesis (two methods with similar bodies should have similar names) does not hold all the time. Generation-based approaches frequently fail also because of two major reasons: the ineffective similarity calculation formula and the immature method name generation techniques. 

We also propose two possible ways for better identifying inconsistent method names. In the future, how to leverage contrastive learning and the ability of LLMs to further improve the performance of identifying inconsistent method names deserves to be investigated.

\section*{Acknowledgments}
The authors would like to say thanks to anonymous reviewers for their insightful comments and suggestions.
This work was partially supported by the National Natural Science Foundation of China [ grant numbers 62172037, 62232003 ].

\section*{Data Availability Statements}
The data that support the findings of this study are openly available in EmpiricalStudy-MCC at https://github.com/Michaelll123/EmpiricalStudy-MCC\\~\citep{MCC}.

\section*{Declarations}
\texttt{Conflict of Interest} The authors declare that they have no conflict of interest.

\bibliographystyle{spbasic}
\bibliography{references}
\end{sloppypar}
\end{document}